\documentclass[aps,prl,reprint,superscriptaddress]{revtex4-2} 
\usepackage[dvipsnames]{xcolor}
\usepackage[colorlinks=true,urlcolor=blue,citecolor=purple,linkcolor=red]{hyperref}
\usepackage{amssymb,amsmath,amsfonts}
\usepackage{epsfig}
\usepackage{graphicx}
\usepackage{epstopdf}
\usepackage{natbib}
\usepackage{braket}
\usepackage[T1]{fontenc}

\def\clock{{\count0=\time
		\divide\count0 60
		\ifnum\count0<10 0\fi\the\count0
		\multiply\count0 -60 \advance\count0 \time
		:\ifnum\count0<10 0\fi \the\count0
}}
\newcommand{\timestamp}{{\small\vbox{\hbox{\tt\jobname.tex}
			\hbox{\the\day/\the\month/\the\year, \clock}}}}

\hypersetup{%
  pdftitle   = {Null matter and the ultrarelativistic origin of hydrodynamics at zero temperature},
  pdfkeywords = {Null fluids, lightlike hydrodynamics, exotic symmetries, geometry},
  pdfauthor  = {Jay Armas, Emil Have, Gianbattista-Piero Nicosia},
  linktoc=page,
}

\newcommand{\C}{\mathbb{C}}

\newcommand{\be}{\begin{eqnarray}}
\newcommand{\ee}{\end{eqnarray}}
\newcommand{\beq}{\begin{eqnarray}}
\newcommand{\eeq}{\end{eqnarray}}

\newcommand{\beqa}{\begin{eqnarray}}
\newcommand{\eeqa}{\end{eqnarray}}
\newcommand{\D}{{\partial}}

\usepackage{mathrsfs}

\newcommand{\so}{\mathfrak{so}}

\newcommand{\MM}{\mathbb{M}}

\newcommand{\NN}{\mathbb{N}}

\newcommand{\verts}[1]{\left\vert #1 \right\vert}

\ifdefined\C
\renewcommand{\C}{\boldsymbol{C}}
\else
\newcommand{\C}{\boldsymbol{C}}
\fi

\renewcommand{\Im}{\operatorname{Im}}

\definecolor{gris}{rgb}{0.5,0.5,0.5}
\definecolor{darkgreen}{rgb}{0.0,0.5,0.0}

\allowdisplaybreaks[0]
\usepackage[scr=boondox]{mathalpha}

\usepackage{etoolbox}
\makeatletter
\pretocmd{\thebibliography}{\let\addcontentsline\@gobblethree}{}{}
\makeatother

\begin{document}

	\title{Null matter and the ultrarelativistic origin of hydrodynamics at zero temperature}

	\author{Jay Armas}
	\email{j.armas@uva.nl}
	\affiliation{Institute for Theoretical Physics, University of Amsterdam, 1090 GL Amsterdam, The Netherlands}
    \affiliation{Dutch Institute for Emergent Phenomena, University of Amsterdam, 1090 GL Amsterdam, The Netherlands}
    \affiliation{Institute for Advanced Study, University of Amsterdam, Oude Turfmarkt 147, 1012 GC Amsterdam, The Netherlands}
    \affiliation{Niels Bohr International Academy, Niels Bohr Institute, University of Copenhagen, Blegdamsvej 17, DK-2100 Copenhagen Ø, Denmark}

	\author{Emil Have}
	\email{emil.have@nbi.ku.dk}
    \affiliation{The Mathematical Institute, University of Oxford, Woodstock Road, Oxford OX2 6GG, United Kingdom}
     \affiliation{Niels Bohr International Academy, Niels Bohr Institute, University of Copenhagen, Blegdamsvej 17, DK-2100 Copenhagen Ø, Denmark}
    \affiliation{Center of Gravity, Niels Bohr Institute, University of Copenhagen, Blegdamsvej 17, DK-2100 Copenhagen Ø, Denmark}
	
	\author{Gianbattista-Piero Nicosia}
    \email{g.nicosia@uva.nl}
    \affiliation{Institute for Theoretical Physics, University of Amsterdam, 1090 GL Amsterdam, The Netherlands}

	\begin{abstract}
   We uncover a universal sector of relativistic fluid dynamics by taking a novel ultrarelativistic limit in which the temperature tends to zero while the flow simultaneously approaches the speed of light. In this regime, hydrodynamics becomes an effective theory of \emph{null matter}, characterised by a preferred null vector, a preferred scale, and their gradients. We show that this theory of null matter constitutes an example of a hydrodynamic theory that can be linearly stable and causal in an arbitrary choice of frame. The framework developed here for null matter can offer insights into ultrarelativistic heavy-ion collisions, astrophysical phenomena with inherently large Lorentz factors, and the dynamics of black hole horizons.
	\end{abstract}

	\maketitle
\textbf{Introduction.} Hydrodynamics is a theory that describes the large-distance, long-time, near-equilibrium behaviour of many systems across a wide range of scales. It is generally valid when the characteristic microscopic length scale, such as the mean free path $\ell_{\text{mfp}}$, is much smaller than the system size $L_{\text{s}}$, or the scale at which the system is being probed, i.e., $\ell_{\text{mfp}}/ L_{\text{s}}\ll1$. In relativistic hydrodynamics, the degrees of freedom are encoded in the thermal vector $\beta^\mu$ arising due to the presence of a preferred thermal rest frame that breaks Lorentz symmetry. The mean free path is typically related to the inverse power of the temperature $T=(-\beta_\mu \beta^\mu)^{-\frac{1}{2}}$. As temperature approaches zero, $\ell_{\text{mfp}}$ tends to diverge and one expects hydrodynamics to break down, the excitations to become ballistic, or to reduce to a trivial theory of dust with no dynamical degrees of freedom (see, e.g., \cite{Lucas:2017idv, 2018arXiv180603933K, Manuel:2004iv, Karch:2008fa, Schmitt:2017efp, Erdmenger:2018svl} for a discussion and instances of $T\sim0$ and diverging mean free paths across various systems with (emergent) Lorentz symmetry).

Despite this expectation, in this letter we show that a novel and well-defined $T\to0$ limit of relativistic hydrodynamics exists in which the fluid velocity $u^\mu=\beta^\mu T$ approaches the speed of light, corresponding to an infinitely boosted velocity. In Minkowski space, where we may parametrise $u^\mu=\gamma U^\mu$ with $\gamma=1/\sqrt{1-\vec v^2/c^2}$ the Lorentz factor, $\vec v^2$ the modulus of the spatial velocity, and $c$ the speed of light, this limit can be achieved by sending $T\to0$ and $\vec v^2\to c^2$ such that $(\varepsilon+\hat P)\gamma^2$ remains finite, where $\varepsilon$ is the energy density and $\hat P$ the pressure of the fluid. Consequently, $\varepsilon+\hat P\to0$ in this limit, suggesting a passing resemblance with certain classes of Carrollian fluids~\cite{deBoer:2021jej, Armas:2023dcz} (and, more generally, framids~\cite{Nicolis:2015sra,Alberte:2020eil}). However, as we will clarify, its origin is of a different nature, arising from the \textit{ultrarelativistic} limit $v^2\to c^2$ rather than the Carrollian limit, where $c\to0$.

Fluids moving at the speed of light are not uncommon. The most well-known example is Bjorken flow, a fluid configuration modelling ultrarelativistic heavy-ion collisions that expands at the speed of light at its boundaries while remaining timelike within the lightcone \cite{PhysRevD.27.140}. In astrophysics, such relativistic fluids partly underpin radiation hydrodynamics, which models the interaction between radiation and matter \cite{1976ApJ...207..244H, Castor_2004}, while in other astrophysical phenomena\textemdash such as active galactic nuclei \cite{2019ApJ...874...43L, 2019ARA&A..57..467B}, gamma-ray bursts \cite{10.1063/1.861619, RevModPhys.76.1143, 2009ApJ...698.1261Z}, and pulsar wind nebulae \cite{1984ApJ...283..710K,Porth:2013boa}\textemdash Lorentz factors can vary between $10$ and $10^6$. In the context of the black hole membrane paradigm \cite{1986bhmp.bookT}, these fluids emerge as an effective description of the intrinsic dynamics of the black hole horizon governed by a subsector of the Einstein equations (see, e.g., \cite{Pinzani-Fokeeva:2014cka}). In general relativity, such fluids model null dust \cite{Bicak:1997bx} with applications to pp-wave spacetimes \cite{Griffiths:1991zp, Blau:2002dy}, gravitational collapse and inflation (e.g., \cite{PhysRevLett.70.13, PhysRevD.57.2279}), as well as in holography (e.g., \cite{Bhattacharyya:2009uu, Chapman:2018dem}), to mention only a few. Yet, we lack a formalism for studying flows close to the speed of light.\newline
\indent As a first step towards laying the groundwork for a more unified description of these diverse physical phenomena, in this letter we systematically formulate relativistic fluid dynamics at the speed of light and show how such special classes of fluids arise as a zero temperature limit of relativistic fluid dynamics. \newline
\indent \textbf{Geometry of null congruences.} Analogously to usual relativistic fluid dynamics formulated on a background $(d+2)$-dimensional spacetime with metric $g_{\mu\nu}$, we consider the existence of a null vector $v^\mu$ satisfying $v^\mu v_\mu=0$ that defines a null congruence. It is useful to introduce an auxiliary null vector $\tau_\mu$ satisfying $\tau^\mu \tau_\mu=0$ and $v^\mu\tau_\mu=-1$, as well as the spatial projector $h_{\mu\nu}=g_{\mu\nu} + 2v_{(\mu}\tau_{\nu)}=:\delta_{AB}e^{A}_\mu e^{B}_\nu$, where $e^A_\mu$ with $A = 1,\dots,d$ are spatial vielbeine satisfying $e^A_\mu v^\mu=0=e^A_\mu \tau^\mu$ and $g^{\mu\nu}e^{A}_\mu e^{B}_\nu=\delta^{AB}$. Demanding that the metric $g_{\mu\nu}$ remains invariant, these objects transform as 
\begin{equation} 
\label{eq:nulltransfinite}
\begin{split}
    v^\mu&\to e^\alpha v^\mu\,,~~~~\tau_\mu\to e^{-\alpha}\Big(\tau_\mu + \Lambda_\mu + \frac{1}{2}\Lambda^2 v_\mu\Big)\,,\\
    e^A_\mu&\to R^A{_B}\left( e^B_\mu+\Lambda^B v_\mu \right)\,,\\
    h_{\mu\nu}&\to h_{\mu\nu} + 2\Lambda_{(\mu}v_{\nu)} + \Lambda^2 v_\mu v_\nu\,,
\end{split}
\end{equation} 
where $R^A{_B}\in \operatorname{O}(d)$ is a spatial rotation, $\Lambda_A$ is the parameter of a null rotation (i.e., a Lorentz boost in the $(\tau_\mu, e_\mu^A)$ plane), and $\alpha$ parametrises a null boost in the $(v^\mu,\tau_\mu)$ plane. In writing the above, we defined $\Lambda^2 := \delta_{AB}\Lambda^A \Lambda^B$ and $\Lambda_\mu = e^A_\mu \Lambda_A$. Below we will use the data in~\eqref{eq:nulltransfinite} to construct the effective theory for null matter, while in the supplementary material more details are given on the origin of these transformations.

\textbf{Effective theory for null matter.} The vector $v^\mu$ does not transform under null rotations and as such Lorentz symmetry is broken in the $(v^\mu, e^A_\mu)$ plane. This is the analogous statement for \emph{null fluids} to the existence of a preferred rest frame that breaks Lorentz symmetry in the context of usual timelike fluids \footnote{\label{fn:null-fluids}The \emph{null fluids} discussed here should not be confused with those introduced in \cite{Banerjee:2015hra}, where it is assumed that the background hosts a non-dynamical null Killing vector field in addition to the timelike Killing vector field that gives rise to temperature and fluid velocity.}. As we are interested in developing an effective theory for null matter coupled to the background metric $g_{\mu\nu}$ and in which $v^\mu$ and, by extension, $\tau_\mu$ are dynamical fields, operators in the theory must be invariant under the various transformations. Inspecting \eqref{eq:nulltransfinite}, there is no invariant structure unless boosts in the $(v^\mu,\tau_\mu)$ plane are broken. Considering first the case of \emph{spontaneously} broken null boosts, we introduce a Goldstone field $\kappa\sim\mathcal{O}(1)$ that transforms under null boosts as $\kappa\to e^{-\alpha}\kappa$, and which allows us to define the null boost-invariant vectors $\ell^\mu=\kappa v^\mu$, $\tilde\ell_\mu=\kappa^{-1}\tau_\mu$ such that $\ell^\mu\tilde\ell_\mu=-1$. It is possible to define ``twisted'' transformations for $\ell_\mu,\tilde\ell_\mu$ akin to \eqref{eq:nulltransfinite} as explained in the supplementary material.

When coupling the theory to $g_{\mu\nu}$, diffeomorphism invariance dictates that the dynamics of null matter is governed by the conservation law
\begin{equation} 
\label{eq:stresscons}
\nabla_\mu T^{\mu\nu}=0\,,
\end{equation}
for the symmetric stress tensor $T^{\mu\nu}$, where $\nabla_\mu$ is the covariant derivative associated with $g_{\mu\nu}$. We require that all local operators describing null matter, such as the stress tensor $T^{\mu\nu}$, are invariant under $\alpha$ and $\lambda_A$ transformations. This implies that such operators can only be functions of the invariant building blocks, namely the dynamical field $\ell^\mu$ and the background metric $g_{\mu\nu}$. 

Following the tenets of low-energy effective field theory, we can provide a general form of $T^{\mu\nu}$ for null matter in a gradient expansion of the fields $\ell_\mu$ and $g_{\mu\nu}$ up to a given order. At zeroth (ideal) order, the stress tensor takes the form 
\begin{equation} 
\label{eq:idealstress}
 T^{\mu\nu}_{(0)}=\mathcal{E} \ell^\mu \ell^\nu+P g^{\mu\nu}\,, \qquad T^{\mu\nu}=T^{\mu\nu}_{(0)}+\mathcal{O}(\D)\,, 
\end{equation}
where both the null energy density $\mathcal{E}=\tilde\ell_\mu\tilde\ell_\nu T^{\mu\nu}$ \footnote{At higher orders in the derivative expansion, the statement $\mathcal{E} = \tilde\ell_\mu\tilde\ell_\nu T^{\mu\nu}$ does not hold and, in particular, is not invariant under null rotations.} and pressure $P=g_{\mu\nu}T^{\mu\nu}/(d+2)$ are constants that specify a particular null fluid. To describe null dust~\cite{Bicak:1997bx}, set $\mathcal{E}=1$ and $P=0$ in~\eqref{eq:idealstress}, leading to an emergent scale invariance, i.e., $g^{\mu\nu}T_{\mu\nu}=0$. Generically we can set $\mathcal{E}=1$ by redefining $\kappa$ as to absorb the constant $\mathcal{E}$.

Plugging the ideal-order stress tensor~\eqref{eq:idealstress} into the conservation equation \eqref{eq:stresscons} leads to 
\begin{equation} \label{eq:compact}
\mathcal{E}a^\mu + \mathcal{E}\ell^\mu \theta=0+\mathcal{O}(\partial^2)\,,
\end{equation}
where we defined the expansion $\theta:=\nabla_\mu\ell^\mu$ and the acceleration $a^\mu:=\ell^\nu\nabla_\nu\ell^\mu$ of the null congruence, and where $\mathcal{O}(\partial^2)$ accounts for gradient corrections to the stress tensor that we will consider later. Projecting \eqref{eq:compact} along $\ell_\mu$ trivially vanishes, while the remaining projections with $\tilde \ell_\mu, h_{\mu\nu}$ yield 
\begin{equation} \label{eq:nullequations}
\theta=\tilde\ell_\mu a^\mu+\mathcal{O}(\partial^2)\,,\qquad a^\mu=c_\ell \ell^\mu+\mathcal{O}(\partial^2)\,,    
\end{equation}
with non-affinity parameter $c_\ell=-\tilde\ell_\mu a^\mu = -\theta$. The freedom associated with $\alpha$ transformations allows us to fix null boosts such that $\theta=0$ (see supplementary material for details), resulting in the dynamics of null geodesics $a^\mu=0+\mathcal{O}(\partial^2)$ at ideal order and with zero expansion ($\theta=0$) at all orders. We note that in $d+2$ dimensions, the vector $\ell^\mu$ has $d+1$ degrees of freedom, while \eqref{eq:stresscons} has $d+2$ equations. The $d$ components of $\tilde\ell_\mu$ may be gauge fixed using $\lambda_A$ transformations. As noted earlier, at ideal order, the projection of \eqref{eq:stresscons} is trivial, eliminating one equation. At higher orders in gradients, this projection should be understood as a constraint equation on null matter.

\textbf{Null matter with a ``preferred scale''.} The total number of equations in \eqref{eq:stresscons} suggests that a more general theory is obtained when null boosts are broken explicitly. In this scenario, besides the existence of a preferred vector $v^\mu$ retaining $d+1$ degrees of freedom, there is also a preferred ``scale'' $\kappa$, neither of which transform under $\alpha$. This implies that the most general stress tensor still takes the form of \eqref{eq:idealstress} with $\ell^\mu\to v^\mu$, but now both the null energy density $\mathcal{E}(\kappa)$ and the pressure $P(\kappa)$ are functions of $\kappa$. The equation of motion \eqref{eq:compact} now becomes
\begin{equation} \label{eq:compact2}
\D^\mu P + v^\mu v^\nu\partial_\nu\mathcal{E} + \mathcal{E}a^\mu + \mathcal{E}v^\mu \theta=0+\mathcal{O}(\partial^2)\,,
\end{equation}
where the acceleration and expansion are defined using $v^\mu$. In this case, the system is fully specified once an equation of state $P(\mathcal{E})$ is provided.  Projecting  \eqref{eq:compact2} along $v^\mu,\tau_\mu, h_{\mu\nu}$ yields
\begin{subequations}
\label{eq:nullequations2}
\begin{align} 
 v^\mu\D_\mu P &=0+ \mathcal{O}(\partial^2) \label{eq:null-eq-a2}\,,\\
 -\nabla_\mu(\mathcal{E}v^\mu)+\mathcal{E}\tau_\mu a^\mu+\tau^\mu \D_\mu P&=0+\mathcal{O}(\partial^2) \,,\label{eq:second-null-eq2}\\
\mathcal{E} h_{\alpha\nu}a^\nu+h_{\alpha}^\mu\D_\mu P&=0+\mathcal{O}(\partial^2)\,.\label{eq:third-null-eq2}
\end{align}
\end{subequations}
Eq.~\eqref{eq:null-eq-a2} expresses that $P$ is conserved along $v_\mu$ and is trivial when $P$ is constant. Eq.~\eqref{eq:second-null-eq2} states that the null momentum current $\mathcal{E}v^\mu$ is not conserved in the presence of pressure gradients and $\tau_\mu a^\mu$ which cannot be gauge-fixed due to the absence of $\alpha$ transformations. The final equation \eqref{eq:third-null-eq2} marks further deviations away from geodesic motion in the presence of spatial gradients of the pressure. Eqs.~\eqref{eq:nullequations2} are invariant under the null rotations in \eqref{eq:nulltransfinite}, that is, if they are satisfied for a specific choice of $\Lambda_A$, then they are satisfied for any other choice. In the supplementary material, an example of a scalar field theory is given in which the pressure $P$ is not constant. A special case of \eqref{eq:compact2} is when $P$ is constant but $\mathcal{E}$ is not. In such circumstances, the stress tensor \eqref{eq:idealstress} acquires a gauge redundancy in which $v^\mu\to \phi(\kappa) v^\mu$ and $\mathcal{E}\to \phi(\kappa)^{-2}\mathcal{E}$ keeps the form of \eqref{eq:idealstress} invariant since $P$ does not transform, reducing the number of degrees of freedom at ideal order to $d+1$. Though similar, this transformation is distinct from the $\alpha$ transformations in \eqref{eq:nulltransfinite} and is not present once gradient corrections are included as we demonstrate in the supplementary material. 

\textbf{First-order corrections.} As is usual in hydrodynamic theory, one may improve the approximation by adding gradient terms up to a given order. This requires the existence of a scale such as $\ell_{\text{mfp}}$ that can be used to control long-wavelength perturbations. For null matter, $\kappa$ provides an effective scale when null boosts are spontaneously broken (e.g., when fixing $c_\ell=0$), or a genuine scale when explicitly broken. With this in mind, we may expand the stress tensor as $T^{\mu\nu}=T^{\mu\nu}_{(0)}+T^{\mu\nu}_{(1)}+\mathcal{O}(\partial^2)$, where $T^{\mu\nu}_{(0)}$ is the ideal order stress tensor in \eqref{eq:idealstress} and $T^{\mu\nu}_{(1)}$ is the correction of order $\mathcal{O}(\partial)$. Using the frame redefinition freedom $\ell_\mu\to\ell_\mu+\delta \ell_\mu$ (or $\kappa\to\kappa+\delta\kappa$ and $v^\mu\to v^\mu+\delta v^\mu$), where $\delta \ell_\mu$ (or $\delta \kappa$ and $\delta v^\mu$) account for at least first-order gradient terms, it is straightforward to show (see supplementary material) that the next-order correction to the stress tensor takes the form
\begin{equation} \label{eq:stress1storder}
T^{\mu\nu}=\mathcal{E} \ell^\mu \ell^\nu+P g^{\mu\nu}+\rho_1\theta g^{\mu\nu}-\eta\sigma^{\mu\nu}+\mathcal{O}(\partial^2)\,, 
\end{equation}
where we have introduced the null shear $\sigma_{\mu\nu}=\nabla_{(\mu}\ell_{\nu)}$ and chosen the analogous frame of the \emph{Landau frame} in timelike fluids. Clearly, $\eta$ is the analogue of shear viscosity in timelike fluids. We have added the term $\rho_1$ for completeness, but it could be removed using the gauge fixing condition $\theta=\mathcal{O}(\partial^2)$. In the explicitly broken case, this term can be removed by a frame transformation. In the case of spontaneously broken boosts, $\rho_1,\eta$ are constant, whereas in the explicitly broken case, $\rho_1(\kappa),\eta(\kappa)$ are functions of $\kappa$, and the expansion/shear is defined using gradients of $v^\mu$ rather than $\ell^\mu$.

\textbf{Low-energy and gapped modes.} The low-energy modes can be obtained by a linearised analysis of \eqref{eq:nullequations} (and of \eqref{eq:nullequations2}). We consider flat Minkowski space $g_{\mu\nu}=\eta_{\mu\nu}$ with coordinates $(t,x^i,z)$ and $i=1,..,d$ and initial configurations with $\ell^\mu=\ell^\mu_0=(1,0^i,\pm 1)$ (or constant $\kappa=\kappa_0$ and $v^\mu=v^\mu_0=(1,0^i,\pm 1)$). These can be thought of as \emph{equilibrium configurations} since $\ell^\mu$ (or $v^\mu$) is a null Killing vector satisfying $\pounds_\ell g_{\mu\nu}=0$ (see supplementary material). We further consider fluctuations $\ell^\mu=\ell_0^\mu+\delta \ell^\mu$ of plane-wave type, i.e., $\delta \ell^\mu(t,x^i,z) = \delta\hat\ell^\mu(\vec k) e^{-i\omega t + ik_ix^i + ik_z z}$ and similarly for $\delta \kappa$ and $\delta v^\mu$. The equations of motion imply the existence of two modes, namely 
\begin{equation} \label{eq:modesgapped}
\begin{split}
&\omega_+=\pm k_z-i\frac{\eta}{2\mathcal{E}}k_i k^i+\mathcal{O}(k^3)\,,\\
&\omega_-=-i\frac{2\mathcal{E}}{\eta}\mp k_z+i\frac{\eta}{2\mathcal{E}}k_i k^i+\mathcal{O}(k^3)\,,
\end{split}
\end{equation}
where $\omega_+$ is a gapless mode and $\omega_-$ is gapped. As expected, $\rho_1$ does not contribute to the modes. In line with hydrodynamics dealing with radiation, the gapless mode $\omega_+$ propagates at the speed of light with attenuation set by the null shear coefficient. Demanding stability requires $\mathcal{E}/\eta>0$, while causality is guaranteed without any need for constraints. Further details are given in the supplementary material together with a treatment of the explicitly broken case, which exhibits the spectrum \eqref{eq:modesgapped} together with a purely advective mode $\omega_0=k_z$ as well as two additional excitations that reduce to $\omega_0$ when $P$ is constant.

\textbf{Lightlike limit of relativistic hydrodynamics}. We will now see exactly how this type of fluid arises as an infinite-boost limit of relativistic fluid dynamics. Consider a relativistic fluid with velocity $u^\mu=\gamma U^\mu$ such that $u^\mu u_\mu=-1$. The limit $\gamma(x)\to \infty$ implements a local infinite boost (see supplementary material for more details). The stress tensor at ideal order takes the perfect fluid form
\begin{equation}\label{eq:stressidealtimelike}
 \hat T^{\mu\nu}=(\varepsilon+\hat P)\gamma^2 U^\mu U^\nu+\hat P g^{\mu\nu}\,,    
\end{equation}
where the energy density $\varepsilon$ and the pressure $\hat P$ are functions of the temperature $T$. The lightlike limit can be taken by sending $\gamma\to\infty$ and $(\varepsilon+\hat P)\to0$ such that
\begin{equation} \label{eq:limitideal}
(\varepsilon+\hat P)\gamma^2\to \mathcal{E}\,,~~~U^\mu\to v^\mu\,,~~~\hat P\to P\,,
\end{equation}
leading to $\hat T^{\mu\nu}\to T^{\mu\nu}$ as in \eqref{eq:idealstress}, where $\mathcal{E}$ is a function of $\kappa$ while $P$ is constant.

The limit \eqref{eq:limitideal} requires that the temperature $T$ approaches a limiting value $T_L$ for which $\varepsilon(T_L)+\hat P(T_L)=0$. In the majority of the examples we encounter, this is precisely the case when $T\to T_L=0$, but, as we shall see, it can also occur for some microscopic theories when $T\to T_L=\infty$. In general, for a neutral fluid, the limit \eqref{eq:limitideal} imposes a particular scaling for the temperature $T\sim \kappa(x)\gamma^a$ for some coefficient $a$ that depends on the microscopic details (see below for examples).

Generically, the scaling of $T$ implies a given scaling for the mean free path $\ell_{\text{mfp}}$ as $\gamma\to\infty$. To ensure that hydrodynamics remains valid, we must therefore scale the length scale at which the system is being probed, i.e., $L_s\sim \gamma^{b}$ for some coefficient $b$ in order to compensate for the diverging mean free path and divergences in the transport coefficients. This has consequences for the limit when departures away from local thermal equilibrium are taken into account. In particular, temporal and spatial scales must be increasingly resolved in the infinite-boost limit $\gamma\to\infty$ such that $\nabla_\mu u_\nu=\nabla_\mu (\gamma U_\nu)\to \gamma^{b+1}\nabla_\mu v_\nu$.

Here, $b$ is a freely choosable exponent arising from scaling the gradients, while the additional power in $b+1$ arises from $u_\nu=\gamma U_\nu$. For the purposes of this work, we take the limit such that $\partial_\mu\log \gamma \to 0$ as $\gamma\to \infty$. In Minkowski space, $\partial_\mu\log \gamma\sim \gamma^2\partial_\mu|\vec v|^2$, implying that the derivatives of the spatial velocity $\vec v$ are suppressed in the lightlike limit. On the other hand, the way $T$ scales with $\gamma$ 
such that $(\varepsilon+\hat P)\to 0$ also fixes the scaling of all transport coefficients appearing at higher orders in the gradient expansion. In order to implement this in a precise manner, and to avoid unwanted divergences, we write the first-order relativistic stress tensor in a non-thermodynamic frame 
\begin{equation} 
\label{eq:LandauSTnon-thermo}
    \hat T^{\mu\nu}_{(1)}=\hat\varsigma \nabla_\alpha u^\alpha g^{\mu\nu}-\hat\eta \nabla^{(\mu}u^{\nu)}\,,   
\end{equation}
where $\hat\varsigma$ is some function of $T$ that involves both bulk and shear viscosity $\hat\eta(T)$. This form can be obtained from the usual Landau frame by a frame transformation as we show in the supplementary material. Using the scaling above for the gradients of $u^\mu$, this implies that as $\gamma\to\infty$ we find
\begin{equation} \label{eq:T1limit}
\hat T^{\mu\nu}_{(1)}
\overset{\gamma\to \infty}{-\hspace{-0.2cm}\longrightarrow}\rho_1 \theta g^{\mu\nu}-\eta\sigma^{\mu\nu}\,,    
\end{equation}
where $\rho_1=\gamma^{b+1}\hat\varsigma$ and $\eta=\gamma^{b+1}\hat\eta$ are  coefficients kept finite in the limit and both functions of $\kappa$, thus obtaining \eqref{eq:stress1storder}. This scaling can also be implemented directly at the level of dispersion relations of timelike fluids by rescaling frequency and wave vector $\omega\to\gamma^{b}\omega$ and $k\to\gamma^{b}k$ thus obtaining the dispersion relations \eqref{eq:modesgapped} in the limit $\gamma\to \infty$ of the shear channel as we show in the supplementary material. We thus demonstrated the existence of a well-behaved lightlike limit of relativistic fluid dynamics.

\textbf{Limits of equations of state and transport.} To show that such limits can also be obtained directly from microscopic theories, we look at equations of state and transport properties obtained from both kinetic theory and holography/gravity. Consider first the thermodynamics of a relativistic gas of massless particles given by
\begin{equation}\label{eq:relmasslessgas}
\begin{split}
&\varepsilon=(d+1)f(d+1)T^{d+2}\,,\qquad\hat P=\frac{\varepsilon}{d+1}\,,\\
&f(d+1)=\frac{2\pi^{(d+1)/2}}{(2\pi)^{d+1}} \frac{\Gamma(d+1)}{\Gamma((d+1)/2)}\,,  
\end{split}
\end{equation}
which are obtained from kinetic theory (see, e.g.,~\cite{Bajec:2024jez}). The limit described above implies that $(d+2)f(d+1)T^{d+2}\gamma^{2}\to \mathcal{E}(\kappa)=(d+2)f(d+1)\kappa^{d+2}$ remains finite as $\gamma\to\infty$ and $\hat P\to P=0$. Thus $T\sim \kappa \gamma^{a}$ with $a=1/(2d+4)$ as $\gamma\to\infty$. Consequently, this limit describes the $T\to0$ limit of this gas of massless particles. 

Another interesting case is the equation of state obtained for a strongly coupled holographic plasma, given by the pressure \cite{Policastro:2001yc, Policastro:2002se, Bhattacharyya:2008mz}
\begin{equation} \label{eq:eqstateholography}
\hat P=\left(\frac{4\pi}{d+2}\right)^{d+1}\frac{T^{d+2}}{4G(d+2)}\,,
\end{equation}
where $G$ is Newton's constant in $d+3$ dimensions and where the pressure satisfies $\varepsilon+\hat P=(d+2)\hat P$ \footnote{We have set the Anti-de Sitter radius $L=1$.}. The lighlike limit requires that $(d+2)\hat P\gamma^2\to\mathcal{E}(\kappa)\sim\kappa^{d+2}$ and $T\sim \kappa \gamma^{1/(2d+4)}$. Similarly, this strongly coupled plasma is characterised by a shear viscosity $\hat \eta/s=1/4\pi$ where $s=\partial\hat P/\partial T$ is the entropy density \cite{Policastro:2001yc}. In the lightlike limit we obtain $\eta=\mathcal{E}^{\frac{d+1}{d+2}}A^{-(d+2)}/(4\pi)$ after choosing $b=d/(d+2)$ and where $A=(4\pi)^{d+1}/((d+2)^{d+1}4G)$. In this case, the lighlike limit describes a $T\to0$ limit of the strongly coupled plasma.

A somewhat different case is that of fluids duals to $p$-dimensional gravitational objects in asymptotically flat spacetime for which the pressure is instead given by \cite{Emparan:2009at}
\begin{equation}
\hat P=-\left(\frac{n}{4\pi}\right)^{n}\frac{T^{-n}}{16\pi G}\,, 
\end{equation}
where $n=d+p+5$ and $G$ is Newton's constant in $d+2$ dimensions. In this case, following the same procedure as above, the lightlike limit implies that $T\to\infty$ and hence the limit describes the high-temperature regime of such fluids.

\textbf{Discussion.} In this letter we have shown that, contrary to expectations, relativistic fluid dynamics admits a well-defined zero temperature limit when the flow velocity approaches the speed of light. This limit is distinct from $T\sim0$ regimes of superfluidity \cite{2000cond.mat..9282G}, condensed matter systems near quantum critical points \cite{PhysRevB.93.075426}, and holographic correlators of black holes near extremality \cite{Edalati:2010pn}. All these cases involve some form of charged fluids, while in this letter we uncovered a universal sector of hydrodynamics, present for any fluid including the simplest uncharged fluids that we focused on.

The systematic formulation of such a limit is expected to be useful not only practically, for instance when considering astrophysical and heavy-ion collision applications with large Lorentz factors, but also conceptually, for example in formulating effective field theories of black hole horizons (see, e.g.,~\cite{Donnay:2019jiz}). These fluids also constitute the starting point for studying timelike fluids expanding at the speed of light at their boundaries and require extending the work of \cite{Armas:2015ssd, Armas:2016xxg}.

Interestingly, these fluids moving at the speed of light provide an example of first-order hydrodynamic theories that can be made linearly stable and causal in any frame as we show in the supplementary material. It is expected that nonlinear stability and causality also holds. Transport coefficients can be extracted by means of Kubo formulae; in particular the null shear viscosity for both the spontaneously and explicitly broken phases can be obtained via
\begin{equation}
\label{eq:Kubo}
\eta=\lim_{\omega\to0}\frac{\Im[G_{T^{xy}T^{xy}}^{R}(\omega,0)]}{-\omega}\,,
\end{equation}
where $G_{T^{xy}T^{xy}}^{R}$ denotes the retarded Green's function. The poles of these Green's functions, analogously to timelike fluids, provide information about the stability properties of the hydrodynamic theory.

Yet another useful application is the study of zero-temperature states in gauge theories via holography. In particular, the limit taken in \eqref{eq:eqstateholography} implies the existence of a gravitational dual to lightlike fluids. In a forthcoming publication, we will show that such duals are non-homogeneous pp-wave geometries due to the presence of higher-derivative corrections~\cite{pp-wave-upcoming}. We also note that all examples of microscopic theories above led to $P=0$ in the limit. However, as already shown in \cite{Emparan:2011hg,Armas:2024dtq}, if the fluid carries a higher-form charge, the pressure is constant and non-zero in the limit. Such effects also leave an imprint on the respective gravitational duals.

The fact that the fluid velocity is null leads to challenges in formulating hydrodynamic effective theories using the same principles as for timelike fluids. In particular, formulating effective actions and equilibrium partition functions requires introducing additional multipliers, similar to actions for massless particles. In addition, the fact that generically the temperature $T\to0$ in this limit suggests that there is no notion of entropy or entropy current associated with these fluids. In the supplementary material, we address some of these questions but further work is required; in particular, it would be interesting to build a Schwinger--Keldysh functional for lightlike fluids along the lines of \cite{Liu:2018kfw}.

Finally, we note that we demonstrated in the supplementary material that the ultrarelativistic limit of the shear channel and the leading $\gamma\to\infty$ behaviour of the sound channel of timelike fluids lead to the spectrum \eqref{eq:modesgapped} of null fluids in the explicitly broken phase. It is likely that fully capturing the ultrarelativistic limit of the sound channel at order $\mathcal{O}(\gamma^{-1})$ requires developing a systematic ultrarelativistic expansion of timelike fluids. We leave this interesting open question for the future.

\textbf{Acknowledgements.}
We are grateful to Matthias Blau, Jan de Boer, Richard Davison, Jos\'e Figueroa-O'Farrill,  Akash Jain, Pavel Kovtun, Ruben Lier, Niels Obers, Natalia Pinzani-Fokeeva, Oliver Porth, and Stefan Vandoren for useful discussions. JA is partly funded by the
Dutch Institute for Emergent Phenomena (DIEP) cluster at the University of Amsterdam via the DIEP programme Foundations and Applications of Emergence (FAEME). The work of EH is supported by Villum Foundation Experiment project 00050317, ``Exploring the wonderland of Carrollian physics'', and Carlsberg
Foundation grant CF24-1656. The Center of Gravity is a Center of Excellence funded by the Danish National Research Foundation under grant No.~184. GPN is partly supported by the Tertiary Education Scholarship Scheme (TESS).

\addtocontents{toc}{\protect\setcounter{tocdepth}{-1}}

\providecommand{\href}[2]{#2}\begingroup\raggedright\endgroup

\addtocontents{toc}{\protect\setcounter{tocdepth}{2}}

\onecolumngrid
\appendix
\setcounter{secnumdepth}{2}

 \renewcommand{\thesection}{\Alph{section}}
 \renewcommand{\thesubsection}{\thesection.\arabic{subsection}}

\makeatletter
 \renewcommand{\appendixname}{Appendix}
 \def\@seccntformat#1{\csname the#1\endcsname\quad}
 \makeatother

\makeatletter
\renewcommand{\p@subsection}{} 
\renewcommand{\thesubsubsection}{\thesubsection.\arabic{subsubsection}}
\makeatletter
\renewcommand{\p@subsubsection}{}
\makeatother

\begin{center}
    \Large\textsc{Supplementary material}
\end{center}

\tableofcontents    
\section{Details on null congruences}
\label{app:null-geom}
In this appendix, we provide additional background on the geometry of null congruences as used in the main text. The study of null congruences was initiated by Raychaudhuri in~\cite{Raychaudhuri:1953yv}, where he introduced his eponymous equation, and then subsequently developed by Sachs~\cite{Sachs:1961zz}, and Sachs, Jordan, and Ehlers~\cite{jordan2013republication}, where they essentially appear in their modern guise. Being a classic subject, there are many references that discuss null congruences: for textbook treatments, see, e.g.,~\cite{Poisson:2009pwt,BlauBook} (and the useful review~\cite{Gourgoulhon:2005ng}), while a detailed discussion of null geodesic congruences in the context of asymptotically flat spacetimes may be found in~\cite{Adamo:2009vu}. A comprehensive and modern treatment of the causal properties of General Relaivity by Witten appears in~\cite{Witten:2019qhl}, while the mathematically inclined reader may find~\cite{Fino:2020uuz} to their liking.

\subsection{General properties of null congruences}
As described in the main text, a null congruence on a $(d+2)$-dimensional Lorentzian manifold $(M,g_{\mu\nu})$ is defined by the set of integral curves of a nowhere-vanishing vector field $v^\mu$, defined up to scale - a rescaling simply leads to a reparametrisation of the integral curves. The triple $(M,g_{\mu\nu},v^\mu)$ defines a Bargmannian manifold~\cite{Figueroa-OFarrill:2020gpr}, with the pair $(g_{\mu\nu},v^\mu)$ forming a Bargmannian structure. If $v^\mu$ is Killing, i.e., 
\begin{equation}
\pounds_v g_{\mu\nu} = 0\,,    
\end{equation}
the Bargmannian structure is equivalent to a Newton--Cartan structure on a $(d+1)$-dimensional manifold via a procedure known as \textit{null reduction}~\cite{Duval:1984cj}. This is the correspondence exploited in~\cite{Banerjee:2015hra} (see also~\cite[App.~A]{Armas:2019gnb}) to construct what they call ``null fluids'', though, as explained in footnote~[{\hypersetup{linkcolor=purple}\hyperref[fn:null-fluids]{30}}], our construction is very different since we consider a \emph{null fluid velocity} rather than an additional background null Killing vector. This means that we consider a fluid moving at the speed of light in the $(d+2)$-dimensional spacetime, and not a Galilean fluid on a $(d+1)$-dimensional Newton--Cartan manifold.

To describe the congruence, it is useful to introduce an auxiliary null vector $\tau^\mu$ satisfying 
\begin{equation}
v^\mu \tau_\mu = -1\,,    
\end{equation}
which allows us to construct the transverse projector $h_{\mu\nu} = g_{\mu\nu} + 2v_{(\mu}\tau_{\nu)}$ which we may express in terms of transverse vielbeine as $h_{\mu\nu} = e^A_\mu e^B_\nu\delta_{AB}$
satisfying $e^A_\mu v^\mu = 0 = e^A_\mu \tau^\mu$. By demanding that the metric remains invariant, the most general allowed transformations of these objects are given in~\eqref{eq:nulltransfinite}, whose infinitesimal version reads
\begin{equation}
\label{eq:infinitesimal-trafos}
    \delta v^\mu = \alpha v^\mu\,,\qquad \delta\tau_\mu = -\alpha\tau_\mu + \lambda_\mu\,,\qquad \delta e^A_\mu =  \lambda^A v_\mu + O^A{_B}e^B_\mu\,,\qquad \delta h_{\mu\nu} = 2\lambda_{(\mu}v_{\nu)}\,,
\end{equation}
with $\lambda_\mu = \lambda_A e^A_\mu$, and where $O^A{_B} \in \so(d)$ is an infinitesimal rotation, i.e., $R^A{_B} = \delta^A_B + O^A{_B}$. Below, in Section~\ref{app:inherited-syms}, we briefly discuss how these symmetries are inherited from the local Lorentz symmetries of the vielbeine if we choose to align one of these with the null vector $v^\mu$. 

Now, if the dual one-form $v = v_\mu dx^\mu$ satisfies the Frobenius condition $v\wedge dv = 0$, the manifold $M$ is foliated by $(d+1)$-dimensional null hypersurfaces, and we remark in passing that it would be interesting to construct a theory of fluids on such null hypersurfaces using the techniques developed in this work. The null congruence is geodesic if 
\begin{equation}
    v^\nu \nabla_\nu v^\mu = c_v v^\mu\,,
\end{equation}
where $c_v$ is a smooth function on $M$ known as the non-affinity. 
\subsection{Inherited symmetries of null congruences}
\label{app:inherited-syms}
The purpose of this subsection is to explicitly identify the transformations~\eqref{eq:nulltransfinite} preserving the frame adapted to the null vector $v^\mu$ as (a subset of) local Lorentz transformations. We do this by decomposing the metric in terms of ``nullbeine'', which are null vielbeine, and then fixing part of the local Lorentz transformations to align one of the nullbeine with $v^\mu$. Such a decomposition is reminiscent of, though not the same as, the one used in the Newman--Penrose formalism~\cite{Newman:1961qr}. Nullbein decompositions are, in particular, very useful for studying the conformal Carrollian structure at null infinity $\mathscr{I}$ in asymptotically flat spacetimes~\cite{Hartong:2025jpp} (see also~\cite{Hartong:2015usd}). 

Consider, as above, a $(d+2)$-dimensional Lorentzian geometry $(M,g)$. In terms of vielbeine (or coframe fields) $\hat e^a_\mu$, where $a=0,1,\dots,d+1$, we may express the metric as 
\begin{equation}
    g_{\mu\nu} = \eta_{ab}\hat e^a_\mu \hat e^b_\nu\,,
\end{equation}
where $\eta_{ab} = \text{diag}(-1,+1,\dots,+1)$ is the $(d+2)$-dimensional Minkowski metric. The vielbeine satisfy the completeness relation $\hat e_\mu^a \hat e^\mu_b = \delta^a_b$, where $\hat e^\mu_b = g^{\mu\nu} \eta_{ba} \hat e^a_\nu$. In turn, we may express the inverse metric $g^{\mu\nu}$ in terms of these inverse vielbeine (or frame fields) as $g^{\mu\nu} = \eta_{ab} \hat e^\mu_a \hat e^\nu_b$. By defining the lightcone combinations
\begin{equation}
    \begin{split}
        \bar U_\mu &= \frac{1}{\sqrt{2}}(\hat e^0_\mu + \hat e^{d+1}_\mu)\,,\qquad \bar V_\mu = \frac{1}{\sqrt{2}}(\hat e^0_\mu - \hat e^{d+1}_\mu)\,, 
    \end{split}
\end{equation}
where both $\bar U$ and $\bar V$ are null, i.e., $\bar U_\mu \bar U_\nu g^{\mu\nu} = 0 = \bar V_\mu \bar V_\nu g^{\mu\nu}$, and which are therefore known as nullbeine, we may recast the metric as
\begin{equation}
    g_{\mu\nu} = -\bar U_\mu \bar V_\nu - \bar V_\mu \bar U_\nu + \delta_{AB}\hat e^A_\mu \hat e^B_\nu\,.
\end{equation}
The vielbeine transform under infinitesimal local Lorentz transformations as
\begin{equation}
    \delta \hat e^a_\mu = \hat \omega^a{_b}\hat e^b_\mu\,.
\end{equation}
Since $\hat \omega$ is antisymmetric, i.e., $\hat \omega_{(ab)} = 0$, it splits as $\hat \omega^a{_b} = \{ \hat \omega^0{_{d+1}},\hat \omega^0{_A},\hat \omega^{d+1}{_A},\hat \omega^A{_B} \}$, leading to the transformations
\begin{equation}
    \begin{split}
        \delta \bar U_\mu &= \alpha \bar U_\mu + \sigma_A \hat e^A_\mu \,,\qquad  \delta \bar V_\mu = -\alpha \bar V_\mu + \lambda_A \hat e^A_\mu \,,\qquad \delta \hat e^A_\mu = O^A{_B}\hat e^B_\mu + \lambda^A U_\mu + \sigma^A V_\mu\,, 
    \end{split}
\end{equation}
where we defined 
\begin{equation}
    \alpha := \hat \omega^0{_{d+1}}\,,\qquad \sigma_A := \frac{1}{\sqrt{2}}(\hat\omega^0{_{A}} + \hat \omega^{d+1}{_{A}}) \,,\qquad \lambda_A := \frac{1}{\sqrt{2}}(\hat\omega^0{_{A}} - \hat \omega^{d+1}{_{A}})\,,\qquad O^A{_B} := \hat \omega^{A}{_{B}}\,.
\end{equation}
Fixing the null rotations parametrised by $\sigma_A$ by aligning $\bar U^\mu$ with $v^\mu$ leads to the transformations in~\eqref{eq:infinitesimal-trafos}, which is the infinitesimal version of~\eqref{eq:nulltransfinite}, leading to the identifications $\bar V_\mu = \tau_\mu$ and $\hat e^A_\mu = e^A_\mu$.

\subsection{Spontaneously vs.~explicitly broken null boosts}
\label{app:broken-null-boosts}
When null boosts are spontaneously broken, the associated Goldstone $\kappa$, which transforms as
\begin{equation}
    \kappa \to e^{-\alpha}\kappa\,,
\end{equation}
allows us construct the null boost-invariant null vector $\ell^\mu=\kappa v^\mu$, and the null boost-invariant auxiliary null vector $\tilde\ell^\mu = \kappa^{-1}\tau^\mu$, though $\tilde\ell^\mu$ still transforms under null rotations. More precisely, the finite transformations in~\eqref{eq:nulltransfinite} now become
\begin{equation} 
\label{eq:nulltransfinite-inv}
\begin{split}
    \ell^\mu&\to \ell^\mu\,,~~~~\tilde\ell_\mu\to \tilde\ell_\mu + \kappa^{-1}\Lambda_\mu + \frac{1}{2}\kappa^{-1}\Lambda^2 v_\mu = \tilde\ell_\mu + \tilde\Lambda_\mu + \frac{1}{2}\tilde\Lambda^2 \ell_\mu \,,\\
    e^A_\mu&\to R^A{_B}\left( e^B_\mu+\Lambda^B v_\mu \right) = R^A{_B}\left( e^B_\mu+\tilde\Lambda^B \ell_\mu \right)\,,\\
    h_{\mu\nu}&\to h_{\mu\nu} + 2\Lambda_{(\mu}v_{\nu)} + \Lambda^2 v_\mu v_\nu = h_{\mu\nu} + 2\tilde\Lambda_{(\mu}\ell_{\nu)} + \tilde\Lambda^2 \ell_\mu \ell_\nu\,,
\end{split}
\end{equation} 
where we defined the parameter of a ``twisted'' null rotation:
\begin{equation}
\label{eq:tilde-Lambda}
    \tilde\Lambda_\mu := \kappa^{-1}\Lambda_\mu\,.
\end{equation}
Infinitesimally, this leads to the transformations (cf.~\eqref{eq:infinitesimal-trafos})
\begin{equation}
\label{eq:infinitesimal-trafos-inv}
    \delta \ell^\mu = 0\,,\qquad \delta\tau_\mu = \tilde\lambda_\mu\,,\qquad \delta e^A_\mu =  \tilde\lambda^A \ell_\mu + O^A{_B}e^B_\mu\,,\qquad \delta h_{\mu\nu} = 2\tilde\lambda_{(\mu}\ell_{\nu)}\,,
\end{equation}
where $\tilde\lambda_\mu$ is the infinitesimal version of the twisted parameter $\tilde\Lambda_\mu$ defined in~\eqref{eq:tilde-Lambda}. Since now $\ell^\mu$ is truly invariant, we may use it to build a null fluid, and decomposing $\nabla_\mu\ell_\nu$ into its symmetric and antisymmetric parts, we get
\begin{equation}
\label{eq:decomp-nabla-ell}
    \nabla_\mu \ell_\nu = \sigma_{\mu\nu} + \omega_{\mu\nu}\,,
\end{equation}
where we have introduced the shear and vorticity of the null congruence according to
\begin{equation}
    \sigma_{\mu\nu} = \nabla_{(\mu}\ell_{\nu)} \,,\qquad \omega_{\mu\nu} = \nabla_{[\mu}\ell_{\nu]}\,,
\end{equation}
and which are both invariant under all local symmetries.  Together with the expansion $\theta=\nabla_\mu \ell^\mu$, this decomposition forms the starting point for the first-order corrections to the null fluid, which we discuss in detail in Section~\ref{app:first-order-corrections}. 

In contrast, when null boosts are explicitly broken due to the existence of a preferred (dynamical) scale (which we again denote by $\kappa$), the null vector $v^\mu$ is invariant under all local transformations; in other words, the null boost transformation with parameter $\alpha$ is absent from the transformations listed in~\eqref{eq:nulltransfinite}. Together, $v^\mu$ and $\kappa$ provide $d+2$ local degrees of freedom, just as in ordinary hydrodynamics. In this case, the decomposition of the invariant first-order tensor $\nabla_\mu v_\nu$ is identical to~\eqref{eq:decomp-nabla-ell}, but with $\ell^\mu \to v^\mu$. The consequences  of explicitly broken null boosts for the fluid description at first order in derivatives are described in more detail in Section~\ref{app:first-order-corrections}.

\section{Details on hydrodynamics at the speed of light}
In this appendix we give a detailed exposition of hydrodynamics at the speed of light. We discuss the notion of equilibrium and first-order corrections in the most general frame, together with details on gauge-fixing conditions. We then perform an exhaustive study of both gapless and gapped modes in an arbitrary frame, and derive conditions on stability and causality. Finally, we show how to extract hydrodynamic correlation functions for null fluids before closing with a brief analysis of a putative null entropy current.

\subsection{``Equilibrium''}
\label{app:equilibrium}
We consider fluids characterised by a null vector coupled to a background metric $g_{\mu\nu}$, whose dynamics can ultimately be derived from an effective action $S$, whose variation takes the form
\begin{equation}
\delta S=\int d^{d+2}x\sqrt{-g}\left(\frac{1}{2}T^{\mu\nu}\delta g_{\mu\nu}+E^\varphi\delta\varphi\right)\,,
\end{equation}
where $T^{\mu\nu}$ is the energy-momentum tensor and $E^\varphi=0$ are the equations of motion for the dynamical fields that we collectively denote by $\varphi$. Constructing the action $S$ is beyond the scope of this work, but its invariance under diffeomorphisms $\delta g_{\mu\nu}=2\nabla_{(\mu}\xi_{\nu)}$ for some vector field $\xi^\mu$ requires that $\nabla_\mu T^{\mu\nu}=0$ when the equations of motion $E^\varphi=0$ are satisfied. At ideal order in gradients we have identified in the letter the stress tensor to be
\begin{equation} \label{eq:stressesideal}
\begin{split}
T^{\mu\nu}_{(0)}=\mathcal{E}\ell^\mu \ell^\nu +P g^{\mu\nu}~&\iff~\text{spontaneously broken null boosts}\,,\\
T^{\mu\nu}_{(0)}=\mathcal{E}(\kappa)v^\mu v^\nu+P(\kappa)g^{\mu\nu}~&\iff~\text{explicitly broken null boosts}\,,
\end{split}
\end{equation}
where in the spontaneously broken case $\ell^\mu=\kappa v^\mu$ with $\kappa\to e^{-\alpha}\kappa$ transforming under null boosts and where the dynamical fields are $\varphi=\{\ell_\mu\}$ or $\varphi=\{\kappa,v^\mu\}$. Working on-shell, $E^\varphi=0$, we are interested in understanding the starting point of any hydrodynamic theory, namely, the notion of ``equilibrium''. In timelike fluids this notion corresponds to the set of time-independent solutions to $\nabla_\mu T^{\mu\nu}=0$. In the case of null fluids we refer to ``equilibrium'' as the set of solutions that are independent along the null direction $\ell_\mu$. We proceed on a case-by-case basis. 

In the spontaneously broken case we assume the existence of a symmetry null vector field $K^\mu$ that acts on the background metric such that
\begin{equation}
\delta_K g_{\mu\nu}=\pounds_K g_{\mu\nu}=0\,,  
\end{equation}
where $\pounds_K$ is the Lie derivative along $K^\mu$. Given that $\pounds_K g_{\mu\nu}=0$ is the Killing equation, $K^\mu$ is a background null Killing vector field. We thus identify $\ell^\mu=K^\mu$ in equilibrium. By definition this implies that $\theta=0$ and because $K^\mu$ is null one finds that 
\begin{equation}
\nabla_\nu(\ell^\mu\ell_\mu)=0\Rightarrow \ell^\mu\nabla_\nu \ell_\mu=\ell^\mu\nabla_\mu \ell_\nu=-a_\nu=0\,,
\end{equation}
where we have used the Killing equation. Thus we see that the conservation law $\nabla_\mu T^{\mu\nu}=\mathcal{E}a^\nu+\mathcal{E}\ell^\nu\theta$ vanishes for such equilibrium configurations. To note is that the Killing equation also implies $\sigma^{\mu\nu}=\nabla^{(\mu}\ell^{\nu)}=0$ in equilibrium and thus that all first order corrections to null fluids vanish. This is analogous to timelike fluids. 

In the explicitly broken case, equilibrium may be achieved via different identifications. In this case, the symmetry parameters are the same, and we now require that
\begin{equation} 
\label{eq:eqconditions}
\delta_K g_{\mu\nu}=\pounds_K g_{\mu\nu}=0\,,\qquad\delta_K \kappa=K^\mu\partial_\mu \kappa=0\,.  
\end{equation}
One can now identify $v^\mu=K^\mu$, which leads to the conservation equation $\nabla_\mu T^{\mu\nu}=\partial^\mu P+v^\mu v^\nu\partial_\nu\mathcal{E}+\mathcal{E}a^\mu+\mathcal{E}v^\mu\theta=h^{\nu}_\mu\partial^\mu P$. We see that the conditions~\eqref{eq:eqconditions} are not sufficient for equilibrium, and that in addition one must impose that spatial gradients of the pressure $h^{\nu}_\mu\partial^\mu P$ vanish. Other identifications are possible, such as $v^\mu=\kappa K^\mu$, but also require the additional condition $h^{\nu}_\mu\partial^\mu P=0$ to be imposed~\footnote{Such ad-hoc conditions on equilibrium typically mean that we lack understanding of the hydrodynamic theory and further work is needed. See \cite{Armas:2018ibg} for a discussion of one such situation in the context of higher-form hydrodynamics.}. When focusing on the limit of timelike fluids, this condition is not important since $P$ is constant. Conditions \eqref{eq:eqconditions} together with $h^{\nu}_\mu\partial^\mu P=0$ are sufficient to set all first-order corrections to zero in equilibrium.

\subsection{First-order corrections}
\label{app:first-order-corrections}
Out of equilibrium, we can proceed as in usual hydrodynamics and promote the symmetry variables to true dynamical fields and correct operators, such as the stress tensor, in a gradient expansion. We provide such a construction in this section.

\paragraph{Spontaneously broken null boosts.}
Addressing first the case of spontaneously broken null boosts, we decompose first-order corrections $T^{\mu\nu}_{(1)}$ as 
\begin{equation} 
\label{eq:stresscorrectionSSB}
T^{\mu\nu}_{(1)}=N g^{\mu\nu}+2\ell^{(\mu}L^{\nu)}+\mathcal{T}^{\mu\nu}\,\iff~\text{spontaneously broken null boosts}\,, 
\end{equation}
where $N$ is a scalar, and both $\ell^{(\mu}L^{\nu)}$ and $\mathcal{T}^{\mu\nu}$ are symmetric traceless structures to be expanded in gradients. We have split the last two structures in \eqref{eq:stresscorrectionSSB} into a term that can be written as the symmetrisation of $\ell^\mu$ and an arbitrary vector $L^\mu$ satisfying $\ell_\mu L^\mu=0$ and a symmetric tensor $\mathcal{T}^{\mu\nu}$ that cannot be written as the symmetrization with $\ell^\mu$. To classify these structures, we use the decomposition \eqref{eq:decomp-nabla-ell} together with the expansion $\theta=g^{\mu\nu}\nabla_\mu \ell_\nu$. Contrary to ordinary timelike fluid dynamics, the shear and the vorticity are not projected transversely to the flow due to the fact that it is not possible to define a projector orthogonal to $\ell^\mu$ that is invariant under null rotations. These allow us to decompose the structures in \eqref{eq:stresscorrectionSSB} as
\begin{equation} \label{eq:stresscorrections}
N=\rho_1\theta\,,\qquad L^\mu= \rho_2\theta\ell^{\mu}+\rho_3a^\mu\,,\qquad \mathcal{T^{\mu\nu}}=-\eta\sigma^{\mu\nu}\,,
\end{equation}
where all coefficients $\rho_{1,2,3}$ and $\eta$ are constant. As in usual hydrodynamics, once gradient corrections are included, ambiguities in the definition of the degrees of freedom $\ell^\mu$ arise due to the redefinition freedom $\ell^\mu\to\ell^\mu+\bar\delta\ell^\mu$ with $\ell_\mu\bar\delta\ell^\mu=0$, where $\bar\delta\ell^\mu$ admit gradient expansions. In turn this freedom implies that $\delta T^{\mu\nu}_{(0)}=2\ell^{(\mu}\bar L^{\nu)}$ where $\bar L^\nu=\bar\delta \ell^\mu$. This freedom actually allows us to choose a frame in which $N=L^\mu=0$. However in the main text we only choose $L^\mu=0$ explicitly leaving $N$ terms in the stress tensor for facilitating the comparison with limits taken of timelike fluids. In summary, the existence of the minimal frame $N=L^\mu=0$ implies that only a single first-order coefficient, namely $\eta$, is needed to characterise a null fluid.

\paragraph{Explicitly broken null boosts.}
We can now apply the same procedure to the case of explicitly broken null boosts. In this context we parameterize the corrections to the stress tensor as in \eqref{eq:stresscorrectionSSB} but with the replacement $\ell^\mu\to v^\mu$ such that
\begin{equation} 
\label{eq:stresscorrectionESB}
T^{\mu\nu}_{(1)}=N g^{\mu\nu}+2v^{(\mu}L^{\nu)}+\mathcal{T}^{\mu\nu}\,\iff~\text{explicitly broken null boosts}\,,
\end{equation}
and where now the structures $N,L^\mu,\mathcal{T}^{\mu\nu}$ are built from $v^\mu, \kappa$ and its gradients. It is straightforward to classify the most general form of these structures
\begin{equation} 
\label{eq:generalframe}
\begin{split}
N=\rho_1\theta+\rho_6 \frac{v^\mu\partial_\mu\kappa}{\kappa}\,,\qquad L^\mu=\rho_2\theta v^{\mu}+\rho_3a^{\mu}+\rho_4v^{\mu}\frac{ v^\alpha\partial_\alpha\kappa}{\kappa}+\rho_5 \frac{\partial^{\mu}\kappa}{\kappa}\,,\qquad \mathcal{T^{\mu\nu}}=-\eta\sigma^{\mu\nu}\,,
\end{split}
\end{equation}
for some arbitrary transport coefficients $\rho_{i}(\kappa),\eta (\kappa)$ with $i=1,..,6$ and $\sigma^{\mu\nu}=\nabla^{(\mu}v^{\nu)}$. We note that we did not explicitly use the ideal-order equations of motion \eqref{eq:nullequations2} to remove some of the terms. Had we done so, using the first equation in \eqref{eq:nullequations2} sets $\chi_P v^\mu \partial_\mu \kappa=\mathcal{O}(\partial^2)$  while \eqref{eq:compact2} allows one to exchange terms of the form $\partial^\mu \kappa$ with terms proportional to $\theta v^\mu$ and $a^\mu$ up to order $\mathcal{O}(\partial^2)$. In any case, field redefinitions act as $\kappa\to\kappa+\bar\delta\kappa$ and $v^\mu\to v^\mu+\bar\delta v^\mu$ with $v_\mu\bar\delta v^\mu=0$, where $\bar\delta\kappa$ and $\bar\delta v^\mu$ admit gradient expansions.  In turn this freedom implies that $\delta T^{\mu\nu}_{(0)}=2v^{(\mu}\bar L^{\nu)}+g^{\mu\nu}\chi_P \delta\kappa$ where $\bar L^\nu=\bar\delta v^\mu+\chi_{\mathcal{E}} v^\mu\bar\delta\kappa$ and we have introduced the analogue of susceptibilities in timelike fluids, namely $\chi_\mathcal{E}=\partial\mathcal{E}/\partial\kappa$ as well as $\chi_P=\partial P/\partial\kappa$. Thus, we can choose a frame in which $\bar\delta\kappa$ and $\bar\delta v^\mu$ are taken such that $N=L^\nu=0$ as in the spontaneously broken case, leaving only one independent coefficient $\eta(\kappa)$. However, in the main text we have also kept $\rho_1$ as it naturally arises from limits of timelike fluids. 

Finally we would like to comment on whether the stress tensor \eqref{eq:generalframe} is invariant under the transformation $v^\mu\to \phi v^\mu$ for some function $\phi(\kappa)$ when $P$ is constant. At ideal order, if $P$ is constant the stress tensor \eqref{eq:stressesideal} acquires the redundancy $\mathcal{E}\to\phi^{-2}(\kappa)\mathcal{E}$ and $v^\mu\to \phi v^\mu$. At first order we perform the transformation in \eqref{eq:generalframe} and find that the following rescalings and shifts are needed for the stress tensor to remain invariant
\begin{equation}
\begin{split}
&\rho_1\to\phi^{-1}\rho_1\,,\qquad \rho_2\to\phi^{-3}\rho_3\,,\qquad \rho_3\to\phi^{-3}\rho_3\,,\qquad \rho_4\to\phi^{-2}\rho_4-\phi^{-3}\phi'\kappa\rho_2-\phi^{-3}\phi' \kappa\rho_3\,,\\
&\rho_5\to\phi^{-1}\rho_5+\frac{\phi^{-2}}{2}\phi'\kappa\eta\,,\qquad \rho_6\to\phi^{-1}\rho_6-\phi^{-2}\phi'\kappa\rho_1\,,\qquad \eta\to\phi^{-1}\eta\,,
\end{split}    
\end{equation}
where $\phi'=\partial\phi/\partial\kappa$. While these transformations make the stress tensor invariant under the change $v^\mu\to\phi v^\mu$, the shifts in coefficients required are indicative of frame transformations. In practice we work with specific stress tensors, say with only $\rho_1$ and $\eta$ coefficients, which transform as 
\begin{equation}
T^{\mu\nu}_{(1)}\to\left(\rho_1\phi\theta+\rho_1v^\mu\partial_\mu\phi\right)g^{\mu\nu}-\eta\phi\sigma^{\mu\nu}-\eta v^{(\mu}\partial^{\nu)}\phi\,.   
\end{equation}
We see that after the transformation the stress tensor acquires gradients of $\phi(\kappa)$ that can only be removed by making a frame transformation $\kappa\to\kappa+\bar\delta\kappa$ and $v^\mu\to v^\mu+\bar\delta v^\mu$. Therefore, in general, the transformation $v^\mu\to\phi v^\mu$ is not an exact redundancy of the stress tensor in any frame. Nevertheless, it is a redundancy of the stress tensor if the focus is only on the strict low-energy regime, in which frame transformations do not affect gapless modes. As is already clear from \eqref{eq:modesgapped}, but also discussed in detail later in this appendix, in order to match the spectrum of null fluids with limits of the spectrum of timelike fluids, we are also interested in the gapped modes for which different frame choices lead to different spectra.

\subsection{Gauge-fixing conditions}
\label{app:gaugefixing}
As we mentioned in the main text, introducing $\tilde \ell^\mu$ (or $\tau_\mu$) and $\kappa$ in the spontaneously broken case adds redundant degrees of freedom. While we do not make use of this gauge fixing in the main letter, here we will first review how to fix such redundancies in the case of null geodesics and then move on to the case of null matter.
\paragraph{Null geodesics.} When dealing with the special case of null geodesics $a^\mu=0$ (affinely parametrised), it is common to fix this freedom by choosing a particular $\kappa$ and a special class of null rotations such that $\ell^\mu \nabla_\mu \tilde \ell^\nu=0$. In particular, given the definition of $\ell_\mu=\kappa v^\mu$ there are $d+1$ degrees of freedom in $\ell_\mu$, but null boosts parametrised by $\alpha$ in \eqref{eq:nulltransfinite} allow us to gauge fix either $\kappa$ or a component of $v^\mu$.  Writing $\ell^\mu \nabla_\mu \tilde \ell^\nu$ out explicitly, we find
\begin{equation}
\ell^\mu\nabla_\mu\tilde\ell_\nu=\kappa v^\mu\nabla_\mu\left(\kappa^{-1}\tau_\nu\right)=v^\mu\nabla_\mu\tau_\nu-\tau_\nu v^\mu\frac{\partial_\mu\kappa}{\kappa} =v^\mu\nabla_\mu\tau_\nu-\tau_\nu v^\mu\partial_\mu \log \kappa\,. 
\end{equation}
Demanding that $\ell^\mu\nabla_\mu\tilde\ell_\nu \overset{!}{=} 0$ amounts to fixing $\kappa$ using null boosts such that
\begin{equation} 
\label{eq:gaugefixing}
v^\mu\nabla_\mu\tau_\nu= c_{\tilde\ell}
\tau_\nu\,,    
\end{equation}
where 
$c_{\tilde\ell}=v^\mu\partial_\mu \log \kappa$. Contracting \eqref{eq:gaugefixing} along $v^\nu$ gives $v^\nu v^\mu\nabla_\mu\tau_\nu=-c_{\tilde\ell}
$. A common choice is to pick a $\kappa$ such that $c_{\tilde\ell}
=0$ leaving only residual constant null boosts. One can check that the gauge fixing condition $c_{\tilde\ell}
=0$ is possible for any particular choice of auxiliary null vector $\tilde\ell^\mu$. We can show this explicitly by looking 
at how the condition $\ell^\nu\ell^\mu\nabla_\mu\tilde\ell_\nu=0$ transforms under (twisted) null rotations with parameter $\tilde\Lambda_\mu = e_\mu^A \tilde\Lambda_A$ (cf.~\eqref{eq:tilde-Lambda}). In particular, using the transformation properties in~\eqref{eq:nulltransfinite-inv}, we find
\begin{equation}  \label{eq:gaugefixphirotation}
\ell^\mu\ell^\nu\nabla_\mu\tilde\ell_\nu \to  \ell^\mu\ell^\nu\nabla_\mu\tilde\ell_\nu+\ell^\mu\ell^\nu\nabla_\mu\left(\tilde\Lambda_\nu+\frac{1}{2}\tilde\Lambda^2\ell_\nu \right)=\ell^\mu\ell^\nu\nabla_\mu\tilde\ell_\nu-\tilde\Lambda_\nu a^\nu=\ell^\mu\ell^\nu\nabla_\mu\tilde\ell_\nu\,,
\end{equation}
where we used that $a^\mu =0$. We thus see that the $\ell^\nu$-projection of the condition $\ell^\mu \nabla_\mu \tilde \ell^\nu=0$ remains invariant under null rotations. On the other hand, fixing the $e^\nu_A$-projection of this condition explicitly breaks null rotations and fixes the $d$ redundant spatial components of $\tilde\ell_\mu$. The transformation of this spatial projection under null rotations reads
\begin{subequations}
\begin{align} 
 e^\nu_A\ell^\mu\nabla_\mu \tilde\ell_\nu&\to  e^\nu_A\left(\ell^\mu\nabla_\mu \tilde\ell_\nu+\ell^\mu\nabla_\mu\tilde\Lambda_\nu\right)+\left(\frac{\tilde\Lambda^2}{2}e^\nu_A-\tilde\Lambda_A \tilde\Lambda^\nu-\tilde\Lambda_A\tilde\ell^\nu\right)a_\nu \label{eq:gaugetransformationspatial}\\
 &= e^\nu_A\left(\ell^\mu\nabla_\mu \tilde\ell_\nu+\ell^\mu\nabla_\mu\tilde\Lambda_\nu\right)\,,
\end{align}
\end{subequations}
where we, once more, used that $a^\mu = 0$. We may now fix null rotations by choosing $\tilde\Lambda_\nu$ such that $e^\nu_A\left(\ell^\mu\nabla_\mu \tilde\ell_\nu+\ell^\mu\nabla_\mu\tilde\Lambda_\nu\right)=0$. This fixes the null frame up to a residual $\tilde\Lambda_\mu=\text{constant}$ transformation.
\paragraph{Null matter.} The details of the gauge fixing procedure are slightly different in the case of null matter because null matter does not necessarily follow geodesic motion. In fact, in the case of spontaneously broken boosts, Eqs.~\eqref{eq:nullequations} state that $a^\mu=\mathcal{O}(\partial^2)$, and hence first-order corrections to the stress tensor violate geodesic motion. Due to this, it is clear from \eqref{eq:gaugefixphirotation} that the gauge-fixing condition for $\kappa$ in the spontaneously broken case adapted to null geodesics (cf.~\eqref{eq:gaugefixing}) is not invariant under null rotations, and in fact it becomes an order-by-order statement in the gradient expansion. Focusing on ideal order, it is possible to gauge fix $\kappa$ by requiring, in analogy with~\eqref{eq:gaugefixing}, that
\begin{equation} 
\mathcal{E}\ell^\mu \ell^\nu\nabla_\mu\tilde\ell_\nu \overset{!}{=} \mathcal{O}(\partial^2)\Rightarrow \mathcal{E}\ell^\mu \ell^\nu\nabla_\mu\tau_\nu=-\mathcal{E}\kappa^2c_{\tilde\ell}+\mathcal{O}(\partial^2)\,,\qquad \text{with} \qquad c_{\tilde\ell}=v^\mu\partial_\mu \log \kappa\,.
\end{equation}
We may now, as above, choose $c_{\tilde\ell}=0$ such that $\mathcal{E}\ell^\mu \ell^\nu\nabla_\mu\tau_\nu=\mathcal{O}(\partial^2)$. Using the equations of motion \eqref{eq:nullequations}, we see that this statement is equivalent to choosing $\kappa$ such that the expansion is subleading in gradients, i.e., $\nabla_\mu\ell^\mu=\theta=\mathcal{O}(\partial^2)$ where we have used that $\mathcal{E}$ is constant. At arbitrarily high order $N$, we can fix $\kappa$ such that the expansion vanishes exactly, i.e., $\theta=0$, which is equivalent to requiring that
\begin{equation} \label{eq:gaugenullmatter}
\mathcal{E}\ell^\mu \ell^\nu\nabla_\mu\tilde\ell_\nu-\tilde\ell_\nu\sum_{i=1}^{N}\nabla_\mu T^{\mu\nu}_{(i)}\overset{!}{=}0\,,
\end{equation}
where $T^{\mu\nu}_{(i)}$ denotes gradient corrections of order $i\in \NN$. It is straightforward to check that this condition is invariant under null rotations using~\eqref{eq:nulltransfinite-inv}:
\begin{equation}
\begin{split}
\mathcal{E}\ell^\mu \ell^\nu\nabla_\mu\tilde\ell_\nu-\tilde\ell_\nu\sum_{i=1}^{N}\nabla_\mu T^{\mu\nu}_{(i)} &\to \mathcal{E}\ell^\mu \ell^\nu\nabla_\mu\tilde\ell_\nu-\tilde\ell_\nu\sum_{i=1}^{N}\nabla_\mu T^{\mu\nu}_{(i)}  -\tilde\Lambda_\nu\left(\mathcal{E}a^\nu + \sum_{i=1}^{N}\nabla_\mu T^{\mu\nu}_{(i)}\right)-\frac{\tilde\Lambda^2}{2}\ell_\nu\sum_{i=1}^{N}\nabla_\mu T^{\mu\nu}_{(i)}\\
&=\mathcal{E}\ell^\mu \ell^\nu\nabla_\mu\tilde\ell_\nu-\tilde\ell_\nu\sum_{i=1}^{N}\nabla_\mu T^{\mu\nu}_{(i)}\,,
\end{split}
\end{equation}
where the last equality follows from the equations of motion $\nabla_\mu T^{\mu\nu}=0$. This shows that the condition \eqref{eq:gaugenullmatter} is the generalisation of the gauge-fixing condition on $\kappa$ in~\eqref{eq:gaugefixphirotation} that we derived for null geodesic congruences. This (on-shell) gauge fixing applies to the case of spontaneously broken null boosts. When null boosts are explicitly broken, $\kappa$ is a genuine degree of freedom and cannot be gauge fixed. However, when $P$ is constant it is possible to redefine $v^\mu$ at ideal order as to impose a gauge-fixing condition similar to $c_{\tilde\ell}=0$.

We may now proceed with gauge fixing null rotations on-shell. It is possible to implement the same canonical choice as for null geodesics, that is, we may fix $\mathcal{E} e^\nu_A\ell^\mu\nabla_\mu \tilde\ell_\nu\overset{!}{=}0$. Using the transformation \eqref{eq:gaugetransformationspatial} under null rotations, this means that we have to account for the presence of a non-zero acceleration $a_\nu$. Using the equations of motion $\nabla_\mu T^{\mu\nu}=0$, this gauge-fixing condition can be recast in the following manner by rewriting the right hand side of \eqref{eq:gaugetransformationspatial} as
\begin{equation} \label{eq:gaugenullmatterspatial}
\mathcal{E}e^\nu_A\left(\ell^\mu\nabla_\mu \tilde\ell_\nu+\ell^\mu\nabla_\mu\tilde\Lambda_\nu\right)-\left(\frac{\tilde\Lambda^2}{2}e^\nu_A-\tilde\Lambda_A \tilde\Lambda^\nu\right)\sum_{i=1}^{N}\nabla_\mu {T^{\mu}}_{\nu(i)}+\tilde\Lambda_A\tilde\ell^\nu\sum_{i=1}^{N}\nabla_\mu {T^{\mu}}_{\nu(i)}\overset{!}{=}0\,,
\end{equation}
where we have included the overall factor of $\mathcal{E}$ and also used the gauge fixing condition \eqref{eq:gaugenullmatter}. While more difficult than the case of null geodesics, it is in principle possible to find a $\tilde\Lambda_\nu$ that satisfies \eqref{eq:gaugenullmatterspatial}. Since the condition \eqref{eq:gaugenullmatterspatial} includes linear terms in $\tilde\Lambda_\mu$, gauge fixing will in general not lead to residual constant-$\tilde\Lambda_\mu$ transformations. We can summarise the gauge-fixing conditions for the case of spontaneously broken null matter in a more succinct way, namely
\begin{equation}
\mathcal{E}\ell^\mu\nabla_\mu\tilde\ell_\nu+\tilde\ell_\nu\tilde\ell_\sigma\sum_{i=1}^{\infty}\nabla_\mu T^{\mu\sigma}_{(i)}\overset{!}{=}0\,, 
\end{equation}
in which we combined \eqref{eq:gaugenullmatter} and $\mathcal{E} e^\nu_A\ell^\mu\nabla_\mu \tilde\ell_\nu=0$. We note that while such a gauge fixing is possible, in the main text we mostly work without introducing $\tilde\ell_\mu$. In the explicitly broken case it is also possible to gauge fix null rotations such that $\mathcal{E} e^\nu_A v^\mu\nabla_\mu \tau_\nu=0$, leading to a slightly more complicated version of \eqref{eq:gaugenullmatterspatial} that also involves gradients of the pressure $P$.

\subsection{Modes}
\label{app:modes}
In this appendix, we compute the modes of the null fluids developed in the main text when $d=2$. We show that when null boosts are spontaneously broken, there are two modes (one gapless and one gapped) with multiplicity 2. The explicitly broken case gives rise to the same modes plus 4 additional ones, one of which is purely advective.

\paragraph{Modes for spontaneously broken null boosts.} In the case of spontaneosuly broken boosts, the energy-momentum tensor is given by~\eqref{eq:stresscorrectionSSB} and we expand around an equilibrium configuration with $\ell_0^\mu =(1,0,0,1)$. The requirement that the combination $\ell_0^\mu + \delta\ell^\mu$ remains null imposes the condition 
\begin{equation}
\label{eq:condition-1}
\delta \ell^t = \delta \ell^z\,,    
\end{equation}
where we defined $\delta\ell^\mu = (\delta\ell^t,\delta \ell^x,\delta \ell^y,\delta \ell^z)$. This means that the perturbed energy-momentum tensor becomes $T^{\mu\nu} = T_0^{\mu\nu} + \delta T^{\mu\nu}$, where
\begin{equation}
    T_0^{\mu\nu} = \mathcal{E} \ell_0^\mu \ell_0^\nu + P \eta^{\mu\nu}\,,\qquad \delta T^{\mu\nu} = 2\mathcal{E}\ell_0^{(\mu}\delta\ell^{\nu)} - \eta \delta \sigma^{\mu\nu} + \rho_1 \eta^{\mu\nu} \delta\theta + 2\rho_2 \ell^\mu_0\ell^\nu_0 \delta\theta + \rho_3\left( \ell_0^\mu \delta a^\nu + \ell_0^\nu\delta a^\mu \right)\,,
\end{equation}
where all coefficients are evaluated in the equilibrium state. The conservation equations are $\D_\mu \delta T^{\mu\nu} = 0$. Projecting with $\ell_0^\mu\eta_{\mu\nu}$ and Fourier transforming, we explicitly get
\begin{equation}
\label{eq:condition-2}
    \ell_{0}^\mu\eta_{\mu\nu}\D_\rho \delta T^{\rho\nu} = 2(k_z - \omega)(\eta - 2 \rho_1)[(k_z - \omega)\delta \ell^t + k_i \delta \ell^i] = 0\,,
\end{equation}
where $i=x,y$. A solution to this equation is $\omega=k_z$ but it can be explicitly checked that for all linearised equations to be solved such solution implies either $k_i=0$ or $k_i\delta\ell^i=0$ and $\eta=0$. Another solution to \eqref{eq:condition-2} requires $\eta = 2\rho_1$ but one can explicitly check that it does not lead to any modes. On the other hand, assuming that $\omega\neq k_z$ and that $\eta \neq 2\rho_1$, we may solve \eqref{eq:condition-2} for $\delta\ell^t$ and find
\begin{equation}
\label{eq:expression-for-ellt}
    \delta \ell^t =  \frac{k_i \delta\ell^i}{\omega - k_z}\,.
\end{equation}
Plugging this into the conservation equations $\D_\mu \delta T^{\mu\nu} = 0$, we find that the combination $\D_\mu \delta T^{\mu x} + \D_\mu \delta T^{\mu y}$ is equivalent to the $t$-component of $\D_\mu \delta T^{\mu\nu}$, itself  a consequence of the identity $k_i\D_\mu\delta T^{\mu i} = (\omega - k_z)\D_\mu \delta T^{\mu t}$, leaving only two independent equations for the two variables $\delta\ell^i$. Writing this system of equations in terms of a matrix acting on the vector $\delta \ell^i$, we find that the requirement that the determinant of this matrix vanishes becomes the condition $F_{\text{shear}}(\omega,k_z,k_i)^2=0$ where we defined the shear polynomial as
\begin{equation}
\label{eq:shearnullpoly}
F_{\text{shear}}(\omega,k_z,k_i)=\left(2\rho_3 +\eta\right)\left(\omega-k_{z}\right)^2+2\left(i \mathcal{E}+\eta k_z\right)\left(\omega-k_{z}\right)-\eta k_i k^i\,.
\end{equation}
The requirement that $F_{\text{shear}}^2$ vanishes gives a fourth-order equation for $\omega$ with two double roots given by
 \begin{equation}
 \label{eq:omega-pm}
     \omega_\pm = -\frac{i\mathcal{E}}{\eta + 2\rho_3} + \frac{2\rho_3 k_z}{\eta + 2\rho_3} \pm \frac{1}{\eta + 2\rho_3}\sqrt{(i\mathcal{E} + \eta k_z )^2 + \eta(\eta + 2\rho_3) k_ik^i}\,,
 \end{equation}
and where we note that $\rho_1$ and $\rho_2$ do not affect the modes. This is consistent with the possible choice of gauge for which $\theta=0$. It is instructive to expand these modes for small $k$, which gives
\begin{equation} \label{eq:omega-pm-smallk}
\omega_+=k_z-i\frac{\eta}{2\mathcal{E}}k_i k^i+\mathcal{O}(k^3)\,,\qquad \omega_-=-i\frac{2\mathcal{E}}{\eta+2\rho_3}-\frac{\eta-2\rho_3}{\eta+2\rho_3}k_z+i\frac{\eta}{2\mathcal{E}}k_i k^i+\mathcal{O}(k^3)\,,
\end{equation}
corresponding to a gapless and a gapped mode. In the case in which $\rho_3=0$, these correspond to the modes in \eqref{eq:modesgapped} while if $\eta=0$ the modes \eqref{eq:omega-pm} truncate to linear order in $k_z$, namely
\begin{equation}\label{eq:modeseta0}
 \omega_+=k_z\,,\qquad \omega_-=-i\frac{\mathcal{E}}{\rho_3}+k_z\,.  
\end{equation}
It is possible to consider a more general equilibrium state of the form $\ell_0^\mu = (\gamma_{\bar v},\bar v,0,1)$ where $\bar v$ is a constant velocity and $\gamma_{\bar v}=\sqrt{1+\bar v^2}$ in which the null vector is slightly tilted in the $x$-direction rather than moving just in the $t-z$-plane. The shear polynomial \eqref{eq:shearnullpoly} is modified to 
\begin{equation} \label{eq:shearpolytilted}
F_{\text{shear}}(\omega,k_{||},k_\perp)=\left(2\rho_3 +\frac{\eta}{\gamma_{\bar v}^2}\right)\left(\frac{\omega-k_{||}}{\gamma_{\bar v}}\right)^2+2\left(i \mathcal{E}+\frac{\eta k_{||}}{\gamma_{\bar v}}\right)\left(\frac{\omega-k_{||}}{\gamma_{\bar v}}\right)-\eta k_\perp^2\,,
\end{equation}
where we have defined $k_{||}=(\bar v k_x+k_z)/\gamma_{\bar v}$ and $k_\perp^2=k^2-k_{||}^2$, as well as $k^2 = k_ik^i + k_z^2$. The modes now become
\begin{equation}
\label{eq:omega-pm-barv}
\omega_\pm
=-\frac{i\mathcal{E}}{(\eta \gamma_{\bar v}^{-1}+2\rho_3\gamma_{\bar v})}+ \frac{2\rho_3\gamma_{\bar v}}{(\eta \gamma_{\bar v}^{-1}+2\rho_3\gamma_{\bar v})}k_\parallel
\pm
\frac{1}{(\eta \gamma_{\bar v}^{-1}+2\rho_3\gamma_{\bar v})}
\sqrt{\left(i\mathcal{E}+\tfrac{\eta}{\gamma_{\bar v}}\,k_\parallel  \right)^2
+ \eta\left(\tfrac{\eta}{\gamma_{\bar v}^2}+2\rho_3\right)\,k_\perp^2} \,,
\end{equation}
and the corresponding small $k$ expansions read
 \begin{align}
\omega_{+} = k_\parallel 
- i\,\frac{\eta}{2\,\mathcal{E}\,\gamma_{\bar v}}\,k_\perp^2 
+ \mathcal{O}(k^3)\,,\qquad  \omega_{-} =
- i\frac{2\mathcal{E}}{(\eta \gamma_{\bar v}^{-1}+2\rho_3\gamma_{\bar v})}
-\frac{\eta(1-\gamma_{\bar v}^2)-2\gamma_{\bar v}^4\rho_3}{\gamma_{\bar v}^2(\eta \gamma_{\bar v}^{-1}+2\rho_3\gamma_{\bar v})}k_\parallel
+ i\,\frac{\eta}{2\,\mathcal{E}\,\gamma_{\bar v}}\,k_\perp^2
+ \mathcal{O}(k^3)\,.
\end{align}
As expected, these expressions reduce to \eqref{eq:omega-pm-smallk} when $\bar v=0$.

\paragraph{Modes for explicitly broken null boosts}
As described in the main text, all quantities depend on $\kappa$ when null boosts are explicitly broken. This means that, while $T_0^{\mu\nu}$ remains the same as in the case of spontanously broken null boosts discussed above, the perturbed energy-momentum tensor now becomes
\begin{equation}
    \delta T^{\mu\nu} = \chi_{\mathcal{E}}  v_0^\mu v_0^\nu \delta\kappa + \chi_{P} \eta^{\mu\nu} \delta\kappa + \rho_1 \eta^{\mu\nu}\delta\theta + 2\rho_2v_0^\mu v_0^\nu \delta\theta + 2\rho_3v_0^{(\mu}\delta a^{\nu)} - \eta_0\delta\sigma^{\mu\nu}\,.
\end{equation}
In this case, the left-hand side of the equations $\D_\mu\delta T^{\mu\nu} = 0$ may be expressed as a matrix acting on the four-component vector $(\delta \kappa, \delta v^i,\delta v^z)$. This matrix has full rank, and its determinant, up to overall factors, can be written as
\begin{equation} \label{eq:generaldeterminant}
(\omega-k_z)F_{\text{shear}}^2F_{\text{long}}\,,    
\end{equation}
where $F_{\text{shear}}$ was given in \eqref{eq:shearnullpoly}, while the longitudinal polynomial $F_{\text{long}}$ takes the form 
\begin{equation}
\label{eq:first-long-poly}
F_{\text{long}}(\omega,k_z,k_i)= \frac{\eta}{2}(\omega^2-k^2)+2i\mathcal{E}(\omega-k_z)+\left(2(\rho_2+\rho_3)+\frac{\chi_{\mathcal{E}}}{\chi_P}\left(\frac{\eta}{2}-\rho_1\right)\right)\left(\omega-k_z\right)^2\,.
\end{equation}
We thus see that the modes $\omega_\pm$ in~\eqref{eq:omega-pm} (each with multiplicity two) arising from $F_{\text{shear}}$ are also present in this case. In addition, the overall factor of $(\omega-k_z)$ gives rise to an advective mode moving at the speed of light, which in the ``rest frame'' $\tilde\omega=\omega-k_z$ corresponds to a zero mode $\tilde\omega=0$, while $F_{\text{long}}$ gives rise to 2 extra modes. We record these modes below
\begin{equation}
\label{eq:omega-pm-ESB}
    \begin{split}
        \omega_0 &= k_z\,,\\
        \tilde\omega_\pm &= \frac{\chi_{\mathcal{E}} k_z(\eta - 2\rho_{1}) + 4\chi_{P} k_z (\rho_{2} + \rho_{3}) - 2i\chi_{P}\mathcal{E}} {\chi_{\mathcal{E}}(\eta - 2\rho_{1}) + \chi_{P} (\eta + 4(\rho_{2} + \rho_{3})) }\\
        &\qquad\pm\frac{\sqrt{\chi_{P} \left[ \chi_{\mathcal{E}}\eta(\eta - 2\rho_{1})k_i k^i + \chi_{P}\left( (2i\mathcal{E} + \eta k_z)^2 + \eta(\eta + 4(\rho_{2} + \rho_{3}) ) k_ik^i\right) \right]  }}{\chi_{\mathcal{E}}(\eta - 2\rho_{1}) + \chi_{P} (\eta + 4(\rho_{2} + \rho_{3})) }\,,
    \end{split}
\end{equation}
where, e.g., $\rho_{1}$ denotes the equilibrium value of $\rho_1(\kappa)$. Expanding the modes $\tilde\omega_\pm$ for small momenta gives
\begin{equation} \label{eq:modesESB}
    \begin{split}
        \tilde\omega_- &= \frac{-4i\mathcal{E} \chi_{P}}{\chi_{\mathcal{E}}(\eta - 2\rho_{1}) + \chi_{P} (\eta + 4(\rho_{2} + \rho_{3}))  } + k_z\left(1 - \frac{2\chi_{P} \eta}{\chi_{\mathcal{E}}(\eta - 2\rho_{1}) + \chi_{P} (\eta + 4(\rho_{2} + \rho_{3})) }\right)+ i\frac{\eta}{4\mathcal{E}}k_ik^i + \mathcal{O}(k^3) \,,\\
        \tilde\omega_+ &= k_z - i\frac{\eta}{4\mathcal{E}}k_ik^i + \mathcal{O}(k^3)\,,
    \end{split}
\end{equation} 
where we explicitly assumed $\chi_P\ne0$. An interesting case is if $\chi_P$ vanishes due to $P$ being constant, as in the limit of timelike fluids discussed in the main letter, or is sub-leading in a suitable $1/\gamma^2$ expansion, as we will discuss in Section \ref{app:detailslimits}. In this situation, in the limit $\chi_P\to0$, the modes in \eqref{eq:omega-pm-ESB} become $\tilde\omega_\pm\to\omega_0= k_z$. For non-zero $\chi_P$, we identify $\tilde\omega_-$ as an additional gapped mode, while $\omega_0$ and $\tilde \omega_+$ are two additional gapless modes (one diffusive and one purely advective). It is possible to also include the terms proportional to 
$\rho_4,\rho_5,\rho_6$ introduced in \eqref{eq:generalframe}, in which case the perturbed energy-momentum tensor includes the following extra terms
\begin{equation}\label{eq:perturbedrho456}
    \delta T^{\mu\nu} \supset 2\rho_4 v_0^\mu v_0^\nu \frac{v^\lambda_0\D_\lambda\delta\kappa}{\kappa} + \frac{\rho_5}{\kappa}( v^\mu_0 \D^\nu\delta\kappa + v^\nu_0 \D^\mu\delta\kappa ) + \rho_6 \frac{v_0^\lambda \D_\lambda \delta\kappa}{\kappa}\eta^{\mu\nu}\,.
\end{equation}
In this case, the determinant still has the same form as in \eqref{eq:generaldeterminant}, with the same $F_{\text{shear}}$ defined in~\eqref{eq:shearnullpoly}, but the longitudinal polynomial is now cubic and involves $\rho_{4,5,6}$ taking the form
\begin{equation}
\label{eq:fullpolylong}
\begin{split}
F_{\text{long}}(\omega,k_z,k_i)=&\left(\frac{\rho_4}{\kappa}(2\rho_1-\eta)-(2\rho_2+\rho_3)\left(\frac{\rho_5+\rho_6}{\kappa}\right)\right)\left(\omega-k_z\right)^3\\
&+i\left(\chi_\mathcal{E}\left(\rho_1-\frac{\eta}{2}\right)-\frac{\mathcal{E}\left(\rho_5+\rho_6\right)}{\kappa}-\chi_P(2\rho_2+\rho_3) \right)(\omega-k_z)^2\\
&+\left(-\frac{F_{\text{shear}}}{2}\left(\frac{\rho_5+\rho_6}{\kappa}\right)-\left(\rho_1-\frac{\eta}{2}\right)\frac{\rho_5}{\kappa}\left(\omega^2-k^2\right)+\chi_P\mathcal{E}\right)\left(\omega-k_z\right)\\
&-i\chi_P\frac{F_{\text{shear}}}{2}\,.
\end{split}
\end{equation}
When $\rho_{4,5,6}$ vanish, this reduces to~\eqref{eq:first-long-poly} up to an overall factor. It is possible to obtain exact expressions (to all orders in $k$) by solving this cubic polynomial, but the expressions are cumbersome. Instead, we report here the expansions in small momenta, namely
\begin{equation}
\label{eq:modesESBfull}
    \begin{split}
        \hat\omega_D &= k_z - i\frac{\eta}{4\mathcal{E}}k_ik^i + \mathcal{O}(k^3)\,,\\
        \hat\omega_\pm&=-i\Gamma_\pm+(1+v_\pm)k_z+\Gamma^\perp_\pm k_i k^i+\Gamma^\parallel_\pm k_z^2+\mathcal{O}(k^3)\,,
\end{split}
\end{equation}     
in which $\hat\omega_\pm$ is a pair of gapped modes and $\hat\omega_D=\tilde\omega_++\mathcal{O}(k^3)$ is a diffusive mode. When $\rho_4=\rho_5=\rho_6=0$ the modes reduce to $\hat\omega_D=\tilde\omega_+$ and $\hat\omega_+=\tilde\omega_-$ given in \eqref{eq:modesESB} while $\hat\omega_-$ is a gapped mode that appears only when $\rho_4,\rho_5,\rho_6\ne0$. We see that in the regime of small momenta $\hat\omega_D$ coincides with $\tilde\omega_+$ and is not affected by $\rho_4,\rho_5,\rho_6$. In \eqref{eq:modesESBfull} we have introduced the damping $\Gamma_\pm$, velocity $v_\pm$ and attenuation coefficients $\Gamma_\pm^\perp, \Gamma_\pm^\parallel$ according to
\begin{equation}
\begin{split}
\Gamma_\pm=&-\frac{\Xi\pm2\sqrt{\Delta}}{4Q}\,,~~v_\pm=\frac{Q_2\Gamma_\pm-\chi_P\eta  }{2Q \Gamma_\pm+\frac{\Xi}{2}}\,,~~\Gamma^\perp_\pm=-i\frac{Q_2\Gamma_\pm-\chi_P\eta}{2\Gamma_\pm(2Q \Gamma_\pm+\frac{\Xi}{2})}\,,~~\Gamma^\parallel_\pm=-i\frac{(3Q \Gamma_\pm+\frac{\Xi}{2})v_\pm^2+2Q_2\Gamma_\pm v_\pm}{\Gamma_\pm(2Q \Gamma_\pm+\frac{\Xi}{2})}\,,
\end{split}
\end{equation}
where we defined
\begin{equation}
\begin{split}
\Xi=&\chi_\mathcal{E}(\eta-2\rho_1)+\chi_P(\eta+4(\rho_2+\rho_3))+\frac{4\mathcal{E}(\rho_5+\rho_6)}{\kappa}\,,\qquad\Delta=4\Xi^2+8\mathcal{E}\chi_PQ\,,\\
Q=&\left(\rho_1-\frac{\eta}{2}\right)\frac{(2\rho_4-\rho_5)}{\kappa}-\frac{(\rho_5+\rho_6)}{\kappa}\left(2\rho_2+2\rho_3+\frac{\eta}{2}\right)\,,\qquad Q_2=\eta\frac{(\rho_5+\rho_6)}{\kappa}+2\left(\rho_1-\frac{\eta}{2}\right)\frac{\rho_5}{\kappa}\,.
\end{split}
\end{equation}
The limit $\rho_4,\rho_5,\rho_6\to0$ of the expressions in \eqref{eq:modesESBfull} must be taken with care but it is possible to show that $\lim_{\rho_4,\rho_5,\rho_6\to0}\hat\omega_+\to\tilde\omega_-$ and $\lim_{\rho_4,\rho_5,\rho_6\to0}\hat\omega_-\to\infty$ as expected for a gapped mode that exists only for $\rho_4,\rho_5,\rho_6\ne0$. In summary, in the general frame \eqref{eq:generalframe} for the explictily broken case, there are 6 modes (3 gapless and 3 gapped). Modes for the more general equilibrium state $\ell^\mu_0=(\gamma_{\bar v}, \bar v, 0,1)$ are straightforward to obtain and boil down to the following substitution $\omega\to\gamma_{\bar v}\omega -\bar v k_x -k_z$ in the polynomials \eqref{eq:shearnullpoly} and \eqref{eq:fullpolylong}.

\subsection{Stability and causality}
In this section we give details on the stability and causality of null fluids in a general frame following the BDNK procedure \cite{Bemfica:2017wps, Kovtun:2019hdm}. In particular we show that any frame can be made stable and causal, by which we mean that the following conditions hold
\begin{equation} \label{eq:stabilitycausality}
 \text {Im}~\omega(k)\le0~\forall~k\,,\qquad \lim_{k\to\infty}\left|\frac{\text{Re}~\omega (k)}{k}\right|\le1\,,\qquad \lim_{k\to\infty}\left|\frac{\text{Im}~\omega (k)}{k}\right|\to0\,.
 \end{equation}
Looking first at the stability condition for the modes \eqref{eq:omega-pm} in the spontaneously broken case in the small $k$ regime \eqref{eq:omega-pm-smallk} requires
\begin{equation}
\frac{\eta}{\mathcal{E}}\ge0\,,\qquad \frac{2\mathcal{E}}{\eta+2\rho_3}>0\,.    
\end{equation}
Generically, the stability of \eqref{eq:omega-pm} also requires that $\eta(\eta+2\rho_3)\ge0$. In turn, at large $k$ we find 
\begin{equation}
\lim_{k\to\infty}\frac{\omega(k)}{k}=\frac{2\rho_3}{\eta+2\rho_3}\bar k\pm\frac{1}{\eta+2\rho_3}\sqrt{\eta^2\bar k^2+\eta(\eta+2\rho_3)(1-\bar k^2)} \,,\qquad \bar k=\frac{k_z}{k}\,. 
\end{equation}
We see that the second causality requirement is satisfied if the stability condition $\eta(\eta+2\rho_3)\ge0$ holds while the first condition imposes in addition that $\rho_3(\eta+2\rho_3)\ge0$. In the special case that $\rho_3=0$, all conditions simply imply that $\eta/\mathcal{E}>0$ whereas in the special case $\eta=0$ they imply $\mathcal{E}/\rho_3>0$. If one assumes future-directed fluids $\mathcal{E}>0$ (the analogue of positive enthalpy in timelike fluids), then all these conditions boil down to
\begin{equation} \label{eq:stabilitySSB}
\eta\ge0\,,\qquad \rho_3\ge0\,.
\end{equation} 
If we consider the more general equilibrium states with $\bar v$ and corresponding modes in \eqref{eq:omega-pm-barv} certain conditions change. In particular the stability of the gap now implies that $\eta \gamma_{\bar v}^{-2}+2\rho_3\ge0$. However, upshot of the analysis is still that \eqref{eq:stabilitySSB} must hold. We conclude that an arbitrary frame composed of $\rho_{1,2,3}, \eta$ can be made stable and causal by imposing \eqref{eq:stabilitySSB}.

In the explicitly broken case the stability and causality conditions derived above for the spontaneously broken case also hold but additional constraints are required for the longitudinal channel to be causal and stable. In particular in the case in which $\rho_{4,5,6}$ vanishes we need 
\begin{equation}
 \frac{\mathcal{E}\chi_P}{\Xi}>0\,,   
\end{equation}
for the gapped mode in \eqref{eq:modesESB} to be stable while stability of $\tilde\omega_+$ is already ensured by the stability of the shear channel. Assuming that $\mathcal{E},\eta>0$ these conditions are sufficient for ensuring stability at all $k$ (for general $\eta$, the condition $\chi_P\eta\Xi>0$ must also be satisfied). On the other hand, for the case in which $\rho_{4,5,6}$ vanishes and the modes are given by \eqref{eq:modesESB}, we find 
\begin{equation}
\begin{split}
\lim_{k\to\infty}\frac{\tilde\omega_\pm}{k}
= &\left(1-\frac{\chi_P \eta}{\Xi}\right)
\bar k\pm
\frac{1}{\Xi}\,
\sqrt{(\chi_P\eta)^2\bar k^2+\chi_P\eta\Xi(1-\bar k^2)}\,.
\end{split}
\end{equation}
Sufficient conditions to ensure reality and boundedness for all $\bar k$, in addition to the stability constraints, are
\begin{equation} \label{eq:causalconditions}
0\le\frac{\chi_P \eta}{\Xi}\le1\,.
\end{equation}
This is possible to satisfy in any frame in which $\rho_{4,5,6}=0$. In the special case in which $\rho_{1,2,3}=0$, assuming $\eta\ge0$, the condition \eqref{eq:causalconditions} holds as long as one demands reasonable conditions on $\chi_P$ and $\chi_\mathcal{E}$, namely $\chi_P,\chi_\mathcal{E}>0$. In the special case in which $\eta,\rho_1=0$, we find $\lim_{k\to\infty}\frac{\tilde\omega_\pm}{k}=\bar k$ and hence there is no need to impose \eqref{eq:causalconditions}. When $\rho_2=\rho_3=0$ conditions \eqref{eq:causalconditions} must be imposed. In summary, when $\rho_{4,5,6}=0$ linear stability and causality can be attained in any frame involving $\eta,\rho_1,\rho_2,\rho_3$.

The general case with $\rho_{4,5,6}\ne0$ is significantly more involved. In particular, stability of the gaps in \eqref{eq:modesESBfull} requires that $\Gamma_\pm\ge0$. In turn, the Routh--Hurwitz criteria applied to \eqref{eq:fullpolylong} leads to the additional conditions
\begin{equation}
Q<0\,,\qquad \Xi>0\,,\qquad \Xi Q_2 +\frac{\eta}{2}\chi_PQ>0\,.   
\end{equation}
These conditions require that $Q<0$ and $Q_2>0$ when taking $\eta,\chi_P>0$. In order to ensure causality we compute the large $k$ limit of the cubic roots. The mode $\hat\omega_D/k$ vanishes at large $k$ while the remaining two take the ballistic form
\begin{equation}
\lim_{k\to\infty}\frac{\hat\omega_\pm}{k}=\frac{(2Q+Q_2)\bar k\pm\sqrt{(Q_2^2+2QQ_2)\bar k^2-2QQ_2}}{2Q}\,.    
\end{equation}
Reality implies $QQ_2<0$, which is already ensured by stability conditions, while boundedness requires that
\begin{equation}
    0\le Q_2\le-2Q\,.
\end{equation}
It can be checked that stability and causality conditions can be ensured in any frame by tuning the coefficients $\rho_{4,5,6}$ appropriately. These conclusions are unaltered in the case of $\ell^\mu_0=(\gamma_{\bar v}, \bar v, 0,1)$. Finally, we note that besides the last two conditions in \eqref{eq:stabilitycausality}, causality also demands that the order $\mathcal{O}_\omega$ of the polynomials from which the dispersion relations are extracted satisfies the following relation 
\begin{equation} \label{eq:finalcausal} \mathcal{O}_\omega\left(F(\omega,k\ne0)\right)=\mathcal{O}_k\left(F(\omega=\mathfrak{d} k,k=\mathfrak{s}^\mu k_\mu)\right)\,,
\end{equation}
where $\mathfrak{d}$ is some real number, $\mathfrak{s}^\mu$ a real unit vector and $F(\omega,k)$ stands for the polynomial in question \cite{Hoult:2023clg}. We can explicitly check that all polynomials \eqref{eq:shearnullpoly}, \eqref{eq:shearpolytilted} and \eqref{eq:fullpolylong} satisfy condition \eqref{eq:finalcausal}. Thus, linear stability and causality can be ensured in all cases of null fluids studied in this work.

\subsection{Correlation functions}
It is customary to study hydrodynamic response functions in timelike fluids. Here we compute response functions using a variational approach (see, e.g., \cite{Kovtun:2012rj}) adapted to null fluids. In the process we recover the same gapless and gapped modes of the previous section as poles in the retarded correlators.

\paragraph{Spontaneously broken null boosts.} The only source that the null fluids considered here couple to is the metric $g_{\mu\nu}$. We thus expand around Minkowski space and an equilibrium configuration for $\ell^\mu=\ell_0^\mu$, namely
\begin{equation}\label{eq:sourcedpert}
    g_{\mu\nu}=\eta_{\mu\nu}+\int\frac{d\omega d^{d}k}{(2\pi)^{d+1}}e^{-i\omega t+ik^{a}x_{a}}\delta g_{\mu\nu}(\omega,k^{a})\ , \ \ \  \ell^{\mu}=\ell^{\mu}_{0}+\delta\ell^{\mu}(\omega,k^{a})\ ,
\end{equation}
where we have Fourier transformed the perturbation in the metric with $a=(i,z)$ and likewise for $\delta\ell^\mu$. We now use the equations of motion for the perturbed stress tensor $\nabla_\mu\left(T^{\mu\nu}+\delta T^{\mu\nu}\right)=0$ in order to obtain the effect on the degrees of freedom due to perturbations in the metric, namely $\delta \ell^\alpha=\delta\ell^\alpha(\delta g_{\mu\nu})$. We will consider the perturbed stress tensor
\begin{equation}
    \delta T^{\mu\nu}=2\mathcal{E}\delta\ell^{(\mu}\ell^{\nu)}_0+P\delta g^{\mu\nu}-\eta \delta\sigma^{\mu\nu}\ + \rho_1 \eta^{\mu\nu} \delta\theta + 2\rho_2 \ell^\mu_0\ell^\nu_0 \delta\theta + 2\rho_3\ell_0^{(\mu} \delta a^{\nu)}\ ,
\end{equation}
where, we have used the null constraint $\pm\delta \ell^z-\delta \ell^t=-(\delta g_{tt}\pm2\delta g_{tz}+\delta g_{zz})/2$ and that $\theta\sim\mathcal{O}(\delta)$ and $a^\mu\sim\mathcal{O}(\delta)$ since we are considering equilibrium configuration states. In particular, we will now focus on equilibrium states of the form $\ell_0^\mu=(1,0,0,1)$. Having solved the perturbed equations of motion for $\delta\ell^\mu$, we can write the linearised stress energy tensor in terms of $\delta g_{\mu\nu}$, which we shall denote by $T^{\mu\nu}_g$. This stress tensor is linear in $\delta g_{\mu\nu}$ but includes all orders of $\omega$ and $k^a$. Defining the retarded Green's function according to
\begin{equation} \label{eq:Greens}
G^{R}_{T^{\alpha\beta}T^{\mu\nu}}=-2\frac{\delta\mathcal{T}_g^{\alpha\beta}}{\delta (\delta g_{\mu\nu})}\Bigg|_{\delta g_{\mu\nu}=0}\,, \qquad \text{with} \qquad \mathcal{T}_g^{\alpha\beta}\equiv\sqrt{-g}T^{\alpha\beta}_{g}\,,
\end{equation} 
we can extract all correlation functions for null fluids. Due to the cumbersome nature of the results, we have opted to show the retarded Green's functions for the particular case when the propagation is along the $z$-direction, i.e., when $k_i=0$. These are given by
\begin{equation} \label{eq:GreensSSB}
\begin{split}
   G_{T^{tt}T^{tt}}^R(\omega,k_z)&=\frac{k_z (2 \mathcal E-i \eta k_z)}{\omega -k_z}-\mathcal E+P\, , \\
   G_{T^{tz}T^{tz}}^R(\omega,k_z)&=-\frac{2 (\mathcal E (k_z+\omega )-\omega  (P+i \eta k_z)+k_z P)}{k_z-\omega }\, ,\\
   G_{T^{zz}T^{zz}}^R(\omega,k_z)&=\frac{k_z (\mathcal E+P)-\omega  (3 \mathcal E+P)+i \eta \omega ^2}{k_z-\omega }\, ,\\
   G_{T^{ti}T^{ti}}^R(\omega,k_z)&=-\frac{i \eta^2 k_z^2}{2 i \mathcal E+\eta \omega +\eta k_z-2 k_z \rho_3+2 \rho_3 \omega }+2 \mathcal E-2 P\, ,\\ 
    G_{T^{zi}T^{zi}}^R(\omega,k_z)&=\frac{2 (\mathcal E+P) (2 \mathcal E-i k_z (\eta-2 \rho_3))-2 i \omega  (\mathcal E+P) (\eta+2 \rho_3)-\eta^2 \omega ^2}{2 \mathcal E-i (\omega  (\eta+2 \rho_3)+k_z (\eta-2 \rho_3))}\, , \\
    G_{T^{ij}T^{kl}}^R(\omega,k_z)&= \delta_{ik}\delta_{jl}\left(2 P+i \eta (k_z-\omega )\right)-\delta_{ij}\delta_{kl}\left(P+\frac{i \eta \rho_1 (k_z-\omega )}{\eta-2 \rho_1}\right)\, ,
\end{split}\end{equation} where one should impose the condition $\eta\neq2\rho_1$ as already seen in Section \ref{app:modes}. The correlators exhibit the presence of two poles located at
\begin{equation}\label{eq:greenpolesSSB}
    \omega_1=k_z\,, \qquad \omega_2 =k_z-\frac{2 (\eta k_z+i \mathcal E)}{\eta+2 \rho_3}\, ,
\end{equation}
which coincide with the expanded modes \eqref{eq:omega-pm-smallk} and \eqref{eq:modesESB} with $k_i=0$ and $k_i=\rho_4=\rho_5=\rho_6=0$, respectively. One can also show that the Green functions with $k_i\neq0$ have poles which are identical to \eqref{eq:omega-pm}.

\paragraph{Explicitly broken null boosts.}
In the explicitly broken case, the procedure employed follows the same steps where now we must also find the dependence of the parameter $\kappa=\kappa_0+\delta\kappa(\omega,k^a)$ on metric fluctuations. In particular from the equations of motion we must extract $\delta \kappa=\delta\kappa(\delta g_{\mu\nu})$. The perturbed stress tensor now takes the form 
\begin{equation}
\label{eq:ESBgreenstress}
    \delta T^{\mu\nu}= (\chi_{\mathcal{E}}v^{\mu}_{0}v^{\nu}_{0}+\chi_{P}\eta^{\mu\nu})\delta\kappa+2\mathcal{E}\delta v^{(\mu}v^{\nu)}_0+P\delta g^{\mu\nu}-\eta\delta\sigma^{\mu\nu}+ \rho_1 \eta^{\mu\nu} \delta\theta + 2\rho_2 v^\mu_0v^\nu_0 \delta\theta + 2\rho_3v_0^{(\mu} \delta a^{\nu)}\,,
\end{equation}
where all parameters $\chi_{\mathcal{E}},\chi_P, \mathcal{E}, P, \eta$ are evaluated on the equilibrium state $\kappa=\kappa_0=\text{constant}$ and where we focused on the case $\rho_{4}=\rho_5=\rho_6=0$. Using \eqref{eq:Greens} we find the correlation functions and once again we will only show the particular case of $k_i=0$. These Green's functions are
\begin{equation}\label{eq:GreensESB}\begin{split}
    G_{T^{tt}T^{tt}}^{R}(\omega,k_z)&=\frac{\mathcal E (\omega -3 k_z)}{k_z-\omega }+P\,,\\ G_{T^{tz}T^{tz}}^{R}(\omega,k_z)&=-\frac{2 \mathcal E (k_z+\omega )}{k_z-\omega }-2 P\,, \\  
    G_{T^{zz}T^{zz}}^{R}(\omega,k_z)&=\mathcal E \left(\frac{2 k_z}{\omega -k_z}+3\right)+P\,,\\ 
    G_{T^{ij}T^{kl}}^{R}(\omega,k_z)&=\delta_{ik}\delta_{jl}\left(2 P+i \eta (k_z-\omega )\right)\\
    &\quad\,+\delta_{ij}\delta_{kl}\Big(\frac{\chi_{\mathcal E} (k_z-\omega ) (-P (\eta-2 \rho_1)-i \eta \rho_1 (k_z-\omega ))}{\chi_{\mathcal E} (\eta-2 \rho_1)
   (k_z-\omega )+\chi_P (-4 i \mathcal E-\omega  (\eta+4 (\rho_2+\rho_3))+k_z (4 (\rho_2+\rho_3)-\eta))}\\  &\quad\,+\frac{\chi_P \left(\mathcal E \eta (k_z-\omega )+4 i \mathcal E P+2 i \eta \rho_2 \omega ^2+2 i \eta
   k_z^2 \rho_2-4 i \eta k_z \rho_2 \omega \right)}{\chi_{\mathcal E} (\eta-2 \rho_1)
   (k_z-\omega )+\chi_P (-4 i \mathcal E-\omega  (\eta+4 (\rho_2+\rho_3))+k_z (4 (\rho_2+\rho_3)-\eta))}  \\
   &\quad\,+\frac{\chi_P \left(k_z P (\eta-4 (\rho_2+\rho_3))+P \omega  (\eta+4 (\rho_2+\rho_3))\right)}{\chi_{\mathcal E} (\eta-2 \rho_1)
   (k_z-\omega )+\chi_P (-4 i \mathcal E-\omega  (\eta+4 (\rho_2+\rho_3))+k_z (4 (\rho_2+\rho_3)-\eta))} \Big)\,,
\end{split}
\end{equation}
while $G_{T^{ti}T^{ti}}^{R}$ and $G_{T^{zi}T^{zi}}^R$ remain the same as those given in \eqref{eq:GreensSSB}. We see that besides the two poles encountered in \eqref{eq:greenpolesSSB} and the $\eta\neq2\rho_1$ condition, in the explicit broken case, we find an additional gapped pole, namely
\begin{equation}\begin{split}
    \omega_{3}&= k_z+\frac{\chi_P (-2 \eta k_z-4 i \mathcal E)}{\chi_{\mathcal E} (\eta-2 \rho_1)+\chi_P (\eta+4 (\rho_2+\rho_3))}\,,    
\end{split}\end{equation}
which coincides with the expanded modes in \eqref{eq:modesESB} when $k_i=0$. If one had to consider the Green functions with $k_i\neq0$, then the poles correspond to \eqref{eq:omega-pm-ESB}. The analysis can be extended to the most general case \eqref{eq:generalframe} in which $\rho_4,\rho_5,\rho_6\ne0$. In this case the perturbations to the stress tensor to be added to \eqref{eq:ESBgreenstress} are identical to \eqref{eq:perturbedrho456}, due to the absence of covariant derivatives in the additional terms. Focusing on the $k_i=0$ case, the retarded Green's functions remain identical to \eqref{eq:GreensESB}, except for the spatial one which is now given by 
\begin{equation}
\begin{split}
    G_{T^{ij}T^{kl}}^{R}(\omega,k_z)&=\delta_{ik}\delta_{jl}\left(2 P+i \eta (k_z-\omega )\right)\\
    &\ \ \ +\frac{\delta_{ij}\delta_{kl}}{A}\Big\{(k_z-\omega ) \Big[\chi_{\mathcal E} \kappa (\eta \rho_1 (k_z-\omega )-i P (\eta-2 \rho_1))+(k_z-\omega) \mathcal E (2 \rho_1 \rho_5-\eta \rho_6)\\ & \ \ \ +(k_z-\omega) P (\eta (2 \rho_4-\rho_6)-2 \rho_1 (2 \rho_4+\rho_5)+4 (\rho_2+\rho_3)(\rho_5+\rho_6))-2k_zP(2\rho_1\rho_5+\eta\rho_6)\\ & \ \ \ -4 i \mathcal E P (\rho_5+\rho_6)+2 i (k_z-\omega)^2\eta \alpha_\rho\Big]+i \chi_P \kappa \Big[\mathcal E \eta (k_z-\omega )+4 i \mathcal E P+2 i \eta \rho_2(k_z-\omega)^2 \\ & \ \ \  +k_z P (\eta-4 (\rho_2+\rho_3))+P \omega  (\eta+4 (\rho_2+\rho_3))\Big]\Big\}\,,
\end{split}
\end{equation} 
and where we have defined $A$ and $\alpha_\rho$ according to  
\begin{equation} 
    A=(\omega-\hat\omega_+)(\omega-\hat\omega_-)|_{k_i=0}\ ,  \ \ \ \alpha_\rho= \rho_1
   \rho_4- \rho_2 \rho_6+2 \frac{\rho_1 \rho_3 \rho_5}{\eta}\ ,
\end{equation} 
with $\hat{\omega}_\pm$ defined in \eqref{eq:modesESBfull}. This immediately implies that the poles are the ones of \eqref{eq:GreensESB} with $\omega_3$ being replaced by the $\hat\omega_\pm$ pair with $k_i=0$. As expected, when considering the case $k_i\neq0$, the (expanded) poles exhibited by the Greens function are identical to \eqref{eq:modesESBfull}. Finally, we note that in general the Green's functions for the explicitly broken phase have the following properties 
\begin{equation}
\begin{split}
    \omega G^{R}_{T^{tt}T^{tt}}(\omega,k^a)-k_{a}G^{R}_{T^{ta}T^{tt}}(\omega,k^a)&=\mathcal{E}(2k_z-\omega)+P\omega\, , \\ 
    \omega G^{R}_{T^{tt}T^{ta}}(\omega,k^a)-k_{b}G^{R}_{T^{tb}T^{ta}}(\omega,k^a)&=k_{a}(P_{0}+\delta_{z,a}\mathcal{E})\, ,\\ 
    \omega k_{z}G^{R}_{T^{tt}T^{za}}(\omega,k^a)-k_{z}k_{b}G^{R}_{T^{tb}T^{za}}(\omega,k^a)&=\delta_{z,a}k_z\omega(P+\mathcal{E})\,, 
\end{split}
\end{equation} 
where the first two relations are reminiscent of timelike fluids perturbed around the rest frame \cite{Kovtun:2012rj} while the last appears due to spatial anisotropies arising from a non-vanishing spatial velocity. Using the Green's functions \eqref{eq:GreensSSB} and \eqref{eq:GreensESB} it is straightforward to extract Kubo formulae such as that given in \eqref{eq:Kubo} in the the main letter.

\subsection{Comments on a putative null entropy current}
In timelike fluids, the existence of a local second law of thermodynamics, which states that divergence of an appropriately defined entropy current is postive semi-definite, $\nabla_\mu S^\mu\ge0$, constrains transport significantly. In the context of null fluids, which we have shown arise as a zero temperature limit of timelike fluids, we expect the entropy, its thermodynamic conjugate variable, to vanish, and hence an entropy current, if it exists, will likely not impose any constraints.  Nevertheless, we entertain this idea in this section by postulating a null entropy current of the form
\begin{equation} \label{eq:entropy}
\begin{split}
S^\mu&=s\ell^\mu-s_1 T^{\mu\nu}_{(1)}\ell_\nu~~\iff~\text{spontaneously broken null boosts}\,,\\
S^\mu&=s(\kappa)v^\mu-s_1(\kappa) T^{\mu\nu}_{(1)}v_\nu~~\iff~\text{explicitly broken null boosts}\,,
\end{split}
\end{equation}
satisfying $\nabla_\mu S^\mu\ge0$, and where in the spontaneously broken case, $s$ and $s_1$ are constants. $S^\mu$ takes the analogue form of timelike fluids in which the first term is the ideal order term while the second is the canonical first order correction to the entropy current. 

Focusing first on the spontaneously broken case, we find
\begin{equation} \label{eq:divergence1}
    \nabla_\mu S^{\mu}=s\theta +\frac{s_1\eta}{2}\ell^\mu\nabla_\mu \theta+\frac{s_1\eta}{2}\theta^2\ge0~\iff~\text{spontaneously broken null boosts}\,,
\end{equation}
where we restricted to the case $N=L^\mu=0$, and used the divergence of \eqref{eq:compact}, namely, $\nabla_\mu a^\mu+\ell^\mu\nabla_\mu\theta+\theta^2=\mathcal{O}(\partial^3)$ to replace some terms. Eq.~\eqref{eq:divergence1} should be analysed order-by-order. At ideal order only the term $s\theta$ appears and in order for the inequality to hold for any configuration of $\ell^\mu$ we must require $s=0$. At first order, the presence of the linear term $\ell^\mu\partial_\mu\theta$ in turn imposes $s_1=0$. Alternatively, we note that the gauge-fixing condition $\theta=0$ described in Section \ref{app:gaugefixing} automatically leads to the vanishing of \eqref{eq:divergence1}. As such, the inequality \eqref{eq:divergence1} is saturated and does not impose any constraints.

Turning our attention to the explicitly broken case, the divergence of the entropy current leads to 
\begin{equation} \label{eq:divergence2}
    \nabla_\mu S^{\mu}=v^\mu\partial_\mu s+ s\theta +\frac{1}{2}a^\mu\partial_\mu \left(s_1 \eta\right)+\frac{s_1\eta}{2}\nabla_\mu a^\mu \ge0~\iff~\text{explicitly broken null boosts}\,.
\end{equation}
Analysing this expression order-by-order in gradients, we see that at ideal order $v^\mu\partial_\mu s\sim\mathcal{O}(\partial^2)$ due to the first equation in \eqref{eq:nullequations2} and hence the second term $s\theta$ requires that $s=0$ since in the explicitly broken case $\theta$ is arbitrary and no gauge-fixing is allowed. At first order in gradients, the term $a^\mu\partial_\mu (\eta s_1)$ can be written as a quadratic term proportional to $a^\mu a_\mu$ using the contraction of \eqref{eq:compact2} with $a^\mu$, namely, $a^\mu\partial_\mu P+\mathcal{E}a^\mu a_\mu=\mathcal{O}(\partial^3)$.  However, the last term in \eqref{eq:divergence2} cannot be written as a purely quadratic term, which can be seen by acting with $\nabla_\mu$ on \eqref{eq:compact2} leading, among other things, to a term linear in $v^\mu\partial_\mu\theta$. Therefore we must require $s_1=0$, resulting in a vanishing entropy current. In summary, postulating \eqref{eq:entropy} does not lead to any constraints. Constraints in null fluids arise due to the requirements of stability and causality, one of which is $\mathcal{E}/\eta>0$ when $N=L^\mu=0$. With the assumption of the null fluid being future-directed $\mathcal{E}>0$ (the analogue of a positive enthalpy in the timelike case), stability requires $\eta>0$ which would be the expected constraint arising from a putative entropy current analysis.

\section{Details on lightlike limits of relativistic fluids and scalar field theories} \label{app:detailslimits}
In this section we give additional details on the lightlike limits of timelike relativistic hydrodynamics. We begin by showing how to take the limit of the ideal order stress tensor and equations of motion. We then move on to include gradient corrections in the two different hydrodynamic frames and show how to implement such limits in their corresponding dispersion relations. At the end we look at scalar field theories with and without dynamical gravity that can be modelled as a relativistic fluid and show how null matter with non-constant pressure and gradient corrections emerges in the limit.

\subsection{The lightlike limit of the fluid velocity as an infinite (local) boost}
\label{app:local-infinite-boost}
Here, we discuss in detail how the lightlike limit of the timelike fluid velocity $u^\mu$ discussed in the main text arises as an infinite Lorentz boost. In $(d+2)$-dimensional Minkowski space, $\MM_{d+2}$, where $d$ denotes the number of spatial directions, and where we may write the metric as $\eta_{\mu\nu} = \text{diag}(-1,1,\dots,1)$, the fluid velocity is a timelike vector $u^\mu$ of the form
\begin{equation}
    u^\mu = \gamma(\vec v) (1,\vec v)^\mu\,,\qquad \gamma(\vec v) = \frac{1}{\sqrt{1-\verts{\vec v}^2}}\,,   
\end{equation}
where $\vec v$ is the $(d+1)$-velocity, and where $\verts{\vec v}$ is its Euclidean norm. By construction, $\eta_{\mu\nu}u^\mu u^\nu = -1$. By infinitely boosting this vector by sending $\verts{\vec v} \to 1$, corresponding to $\gamma(\vec v) \to \infty$, the vector $U^\mu = (1,\vec v)^\mu$ becomes null, i.e., $\eta_{\mu\nu}U^\mu U^\nu \to 0$ as $\verts{\vec v} \to 1$. Alternatively, we can describe this procedure by explicitly exhibiting the infinite boost transformation that gives rise to the infinite $\gamma$-factor. To this end, choose a spacelike unit vector $ n^\mu = (0,\vec n)^\mu$ and decompose $u^\mu$ in lightcone components relative to $n^\mu$, i.e.,
\begin{equation}
\label{eq:LC-split}
    u^\pm = u^0 \pm \eta_{\mu\nu}u^\mu n^\nu\,,\qquad \eta_{\mu\nu}n^\mu n^\nu = +1\,.
\end{equation}
A boost with rapidity $\zeta = \tanh^{-1}\verts{\vec v} $ in the $n^\mu$-direction (for example, $n = \D_z$) acts diagonally on lightcone components according to
\begin{equation}
    u^\pm \mapsto e^{\mp\zeta} u^\pm\,,
\end{equation}
while the remaining transverse components remain inert. Writing $u = u^+ \D_+ + u^- \D_- + u^\perp$, where $\D_\pm$ are the lightcone components corresponding to the split in~\eqref{eq:LC-split}, the limit $\zeta \to \infty$ gives $u\mapsto e^\zeta u^-\D_-$, or, in Cartesian components,
\begin{equation}
    u^\mu \, \overset{\zeta\to \infty}{\scalebox{1.8}{$\sim$}}\, \frac{e^\zeta }{2}(1,\vec n)^\mu\,,
\end{equation}
where $(1,\vec n)$ is a null vector. 

How does this generalise to curved spacetime? Let $(M,g)$ be a Lorentzian manifold with metric $g_{\mu\nu}$, and let $u^\mu$ be a (normalised) timelike vector satisfying $g_{\mu\nu}u^\mu u^\nu = -1$. We can write the (inverse) metric in terms of inverse vielbeins (or frame fields) $e^\mu_a$ for $a =0,\dots,d$ as $g^{\mu\nu} = \eta^{ab} e^\mu_a e^\nu_b$, and, in particular, we may decompose the fluid velocity as
\begin{equation}
    u^\mu = u^a e_a^\mu\,.
\end{equation}
At every point $p\in M$, we then perform a local boost~\footnote{This local Lorentz transformation is an \emph{active} transformation that changes the vector and should not be confused with the (passive) local Lorentz transformations that make up a gauge redundancy of the frame fields.} in the (say) $e_{d+1}^\mu$-direction, i.e.,
\begin{equation}
    u^a \mapsto u'^a\,,\qquad u'^0 = u^0\cosh \zeta  - u^{d+1}\sinh \zeta \,,\qquad u'^{d+1} = u^{d+1}\cosh\zeta - u^0 \sinh\zeta\,, \qquad u'^a = u^a ~~\text{for}~~a\neq 0,d+1\,,
\end{equation}
where $\zeta$ is now the rapidity of a local Lorentz boost. Under an infinite local Lorentz boost, we now have
\begin{equation}
    u^\mu \, \overset{\zeta\to \infty}{\scalebox{1.8}{$\sim$}}\, \frac{e^\zeta }{2} (e^\mu_0 - e^\mu_{d+1})\,,
\end{equation}
where $(e^\mu_0 - e^\mu_{d+1})$ is a null vector.

\subsection{Ideal-order lightlike limit of relativistic fluids}
We will first consider in detail the limit of ideal-order relativistic hydrodynamics.  The stress tensor of an ideal perfect fluid takes the form
\begin{equation}
\hat T^{\mu\nu}_{(0)}=(\varepsilon+\hat P)u^\mu u^\nu + \hat P g^{\mu\nu} \,,
\end{equation}
where we parametrise $u^\mu=\gamma U^\mu$ in which $U^\mu = U^\mu(x,\gamma(x))$ is a timelike vector with norm $U^\mu U_\mu=-1/\gamma^2$. We thus see that in the ultrarelativistic limit, corresponding to an infinitely boosted velocity, where $\gamma\to\infty$, we have that $\lim_{\gamma\to\infty}U^\mu U_\mu\to0$, and $U^\mu\to v^\mu$ becomes a null vector. The limit we described in the main letter is taken such that the stress tensor remains finite in this limit. Defining the enthalpy $w=\varepsilon+\hat P$, the avoidance of divergences requires \footnote{Limits can also be taken of timelike equilibrium configurations for which $T=\gamma T_0$ for some constant temperature $T_0$. In this case, we must scale $T_0$ with an appropriate power of $\gamma$ such that $w\gamma^2$ is kept finite.}
\begin{equation} 
\label{eq:idealorderlimit}
w\gamma^2\to\mathcal{E}(\kappa) \,,\qquad U^\mu\to v^\mu\,. 
\end{equation}
Generically the limit is taken such that $w\to f(x)/\gamma^{2}+\mathcal{O}(\gamma^{-3})$ for some function $f(x)$ which we identify as $\mathcal{E}(\kappa)=f(\kappa(x))$, in which we took into account that all thermodynamic quantities for a neutral fluid are functions of $T$. In particular, the limit \eqref{eq:idealorderlimit} implies a particular scaling for the temperature $T$ that is dependent on the microscopic details, and which we parametrize as $T\to \kappa(x) \gamma^a$ for some coefficient $a$. This implies that in the limit $\gamma\to\infty$, all scalar functions are functions of $\kappa$.

On the other hand, we assume that  \eqref{eq:idealorderlimit} implies $\hat P=P+P_1(x)/\gamma^2+\mathcal{O}(\gamma^{-3})$ where $P$ is a constant and $P_1(x)$  is some function on spacetime \footnote{It is conceivable that there could be a term of order $1/\gamma^{b_1}$ in the expansion of $\hat P$, where $0<b_1<2$, but we do not consider this possibility here.}. Thus in the limit, we find
\begin{equation}
\lim_{\gamma\to\infty}\hat T^{\mu\nu}_{(0)} \to T^{\mu\nu}_{(0)}=\mathcal{E}(\kappa)v^\mu v^\nu+P g^{\mu\nu}\,, 
\end{equation}
where $P$ is constant and $\mathcal{E}(\kappa)$ is a function on spacetime. It is interesting to understand how the limit of the timelike relativistic equations of motion $\nabla_\mu \hat T^{\mu\nu}_{(0)}=0$ gives rise to the null equations of motion $\nabla_\mu T^{\mu\nu}_{(0)}=0$. Because the equations of motion naturally involve gradients of the fluid variables, we need to specify their limiting value. In particular, we assume that gradients of fluid variables behave as 
\begin{equation} 
\label{eq:limitgrad}
\lim_{\gamma\to\infty}\frac{1}{\gamma}\nabla_\mu u_\nu = \nabla_\mu v_\nu\,,
\end{equation}
and hence remain finite in the limit. In general we are assuming that in the limit
\begin{equation}
\label{eq:log-vanishes}
    \D_\mu \log \gamma \to 0\qquad \text{as}\qquad \gamma\to\infty\,.
\end{equation}
Given this, we write the fluid projection $u_\nu\nabla_\mu \hat T^{\mu\nu}_{(0)}=0$ as 
\begin{equation} \label{eq:limituprojection}
\begin{split}
u_\nu\nabla_\mu \hat T^{\mu\nu}_{(0)}&=-\nabla_\mu(w u^\mu)+u^\mu\partial_\mu \hat P=-\nabla_\mu\left(\frac{w\gamma^2 U^\mu}{\gamma}\right)+\gamma U^\mu \partial_\mu\left(\frac{P_1}{\gamma^2}\right)\\
&=\frac{1}{\gamma}\left[-\nabla_\mu\left(w\gamma^2 U^\mu\right)+w\gamma^2U^\mu\partial_\mu \log  \gamma+\gamma^2 U^\mu \partial_\mu\left(\frac{P_1}{\gamma^2}\right)\right]=0\,,\\
&\overset{\gamma\to \infty}{-\hspace{-0.2cm}\longrightarrow}~\frac{1}{\gamma}\left[-\nabla_\mu\left(\mathcal{E}v^\mu\right)+v^\mu\partial_\mu P_1\right]=0\,,
\end{split}
\end{equation} 
where we used \eqref{eq:idealorderlimit} and \eqref{eq:limitgrad} in the last line. We can perform a similar exercise starting with the projection $P_\alpha^\nu\nabla_\mu \hat T^{\mu\alpha}_{(0)}=0$, where $P^{\mu\nu}=g^{\mu\nu}+u^\mu u^\nu$ is the orthogonal projector to $u^\mu$. We find
\begin{equation}\label{eq:limitSprojection}
\begin{split}
P_\alpha^\nu\nabla_\mu \hat T^{\mu\alpha}_{(0)}&=w \hat a^\nu +\partial^\nu P+ u^\nu u^\mu\partial_\mu P\\
&=w\gamma^2\left(U^\mu\nabla_\mu U^\nu+U^\nu U^\mu\partial_\mu\log \gamma\right)+\partial^\nu\left(\frac{P_1}{\gamma^2}\right)+\gamma^2U^\nu U^\mu\partial_\mu\left(\frac{P_1}{\gamma^2}\right)=0\,,\\
&\overset{\gamma\to \infty}{-\hspace{-0.2cm}\longrightarrow}~\mathcal{E}a^\nu+v^\nu v^\mu\partial_\mu P_1+\mathcal{O}(\gamma^{-2})=0\,,
\end{split}
\end{equation} 
where we have used the definition $\hat a^\nu=u^\mu\nabla_\mu u^\nu$. We can now substitute \eqref{eq:limituprojection} into \eqref{eq:limitSprojection} to find
\begin{equation} 
\label{eq:idealorderlimit-app}
v^\nu\nabla_\mu\left(\mathcal{E}v^\mu\right)+\mathcal{E}a^\nu=0\,,
\end{equation}
which is precisely \eqref{eq:compact2} for constant pressure $P$. Contracting the limit in \eqref{eq:limitSprojection} with the auxiliary vector $\tau_\nu$ we find $\mathcal{E}\tau_\nu a^\nu=v^\mu\partial_\mu P_1$, which when used in \eqref{eq:limituprojection} leads to
\begin{equation}
-\nabla_\mu\left(\mathcal{E}v^\mu\right)+\mathcal{E}\tau_\nu a^\nu=0\,,    
\end{equation}
which is identical to the second equation in \eqref{eq:nullequations2} when $P$ is constant. We have thus recovered the equations for null fluids in the explicitly broken phase from a limit of the timelike fluid equations. We note in particular that the condition $\mathcal{E}\tau_\nu a^\nu=v^\mu\partial_\mu P_1$ derived from \eqref{eq:limitSprojection} is indicative of the explicitly broken phase since we cannot simultaneously rescale $v^\mu \to\ell^\mu=\sqrt{\mathcal{E}} v^\mu$ and set $\tau_\mu a^\nu=0$ by gauge fixing, contrary to the spontaneously broken case. At the same time, the fact that the pressure $P$ is constant at leading order implies that temperature fluctuations are suppressed in the limit.

\subsection{First-order lightlike limits of relativistic fluids} \label{app:firstorderlimits}
The limit of the stress tensor and the derivation of the corresponding equations of motion proceed along the same lines as at ideal order. We begin with the first-order corrections written in Landau frame
\begin{equation} \label{eq:LandauST}
    \hat T^{\mu\nu}_{(1)}=-\hat\zeta\hat \theta P^{\mu\nu}-\hat\eta \hat\sigma^{\mu\nu}\,,
\end{equation}
where $\hat\theta=\nabla_\mu u^\mu$, $\hat \sigma^{\mu\nu}=P^{\mu\alpha}P^{\nu\sigma}\nabla_{(\alpha}u_{\sigma)}-\hat\theta P^{\mu\nu}/(d+1)$ and $\hat\zeta,\hat\eta$ are bulk and shear viscosities, respectively; both functions of $T$. Before taking limits it is useful to write \eqref{eq:LandauST} in the form 
\begin{equation} 
\label{eq:LandauST2}
\begin{split}
\hat T^{\mu\nu}_{(1)}&=\tilde\zeta\hat \theta u^\mu u^\nu  +\tilde\zeta\hat \theta g^{\mu\nu}-\hat\eta\nabla^{(\mu} u^{\nu)}-\hat\eta u^{(\mu}\hat a^{\nu)}\\
&=\tilde\zeta\gamma^3\left(\nabla_\mu U^\mu+U^\mu\partial_\mu\log \gamma\right)U^\mu U^\nu+\tilde\zeta\gamma \left(\nabla_\mu U^\mu+U^\mu\partial_\mu\log \gamma\right)-\hat\eta\gamma\left(\nabla^{(\mu}U^{\nu)}+U^{(\mu}\partial^{\nu)}\log \gamma\right)\\
&-\hat\eta\gamma^3\left(U^{(\mu}U^{\alpha}\nabla_\alpha U^{\nu)}+U^{(\mu}U^{\nu)}U^\alpha\partial_\alpha\log \gamma\right)\,,
\end{split}
\end{equation}
where $\tilde\zeta=-\hat\zeta +\eta/(d+1)$. We note that the first and last terms in \eqref{eq:LandauST2} naturally diverge as $\gamma^{3}$ as $\gamma\to\infty$, while the remaining two diverge with the slower rate $\gamma$. This suggests that $\hat\eta\gamma^3$ and $\tilde\zeta\gamma^3$ should be kept finite in the limit. However, given that both $\hat\eta$ and $\hat\zeta$ are functions of $T$ and the scaling of $T$ is fixed by \eqref{eq:idealorderlimit-app}, the scalings $\hat\eta\gamma^3$ and $\tilde\zeta\gamma^3$ would only give finite results for very particular cases of equations of state and in particular spacetime dimensions. Instead, we note that gradients are characterised by a length scale $L_s$ of local perturbations, that is $\mathcal{O}(\partial)\sim L_s$, and we can choose to scale the gradients with an appropriate power of $\gamma^b$, i.e., $L_s\sim\gamma^{b}$. This implies that $\nabla_\mu\to\gamma^{b}\nabla_\mu$ as $\gamma\to\infty$, where $b$ is chosen such that particular coefficients remain finite. This does not affect the analysis at ideal order, however expressions such as \eqref{eq:limitSprojection} get rescaled by a factor of $\gamma^b$. 

The rescaling of gradients suggests two different ways of taking the limit $\gamma\to\infty$. Namely, it is possible to take the limit directly in \eqref{eq:LandauST2}, in which case the first and last terms remain finite in the limit, while the remaining vanish. Alternatively, a frame transformation to a non-thermodynamic frame (see Eq.~\eqref{eq:LandauSTnon-thermo}) can be performed, in which case the first and third terms in \eqref{eq:LandauST2} remain finite. Here we explore both possibilities.
\paragraph{Landau frame.}
Considering first the Landau frame \eqref{eq:LandauST2}, and performing the limit directly by scaling the gradients and using~\eqref{eq:log-vanishes}, we find 
\begin{equation} 
\label{eq:T1limit2}
\hat T^{\mu\nu}_{(1)}
\overset{\gamma\to \infty}{-\hspace{-0.2cm}\longrightarrow}2\rho_2\theta \ell^\mu \ell^\nu + 2\rho_3 \ell^{(\mu}a^{\nu)}\,, 
\end{equation}
where $2\rho_2=\gamma^{b+3}\tilde\zeta$ and $2\rho_3=-\gamma^{b+3}\hat\eta$ remain finite in the limit. This form of the stress tensor agrees with the general form in \eqref{eq:stresscorrections}. We did not focus on this case in the core of the letter because both terms in \eqref{eq:T1limit2} can be removed using the redefinition freedom $\ell^\mu\to\ell^\mu+\delta\ell^\mu$. The limit can also be taken at the level of the equations of motion. Consider the $u_\mu$ projection of the equation of motion
\begin{equation} \label{eq:limituprojection2}
\begin{split}
u_\nu\nabla_\mu \hat T^{\mu\nu}&=-\nabla_\mu(w u^\mu)+u^\mu\partial_\mu \hat P+\hat\zeta\hat\theta^2+\hat\eta \hat\sigma^{\mu\nu}\hat\sigma_{\mu\nu}\\
&=\frac{1}{\gamma}\left[-\nabla_\mu\left(w\gamma^2 U^\mu\right)+w\gamma^2U^\mu\partial_\mu \log  \gamma+\gamma^2 U^\mu \partial_\mu\left(\frac{P_1}{\gamma^2}\right)+\tilde\zeta\gamma^{b+3}\left(\nabla_\mu U^\mu+U^\mu\partial_\mu\log \gamma\right)^2\right]=0\,,\\
&\overset{\gamma\to \infty}{-\hspace{-0.2cm}\longrightarrow}~\frac{1}{\gamma}\left[-\nabla_\mu\left(\mathcal{E}v^\mu\right)+v^\mu\partial_\mu P_1-2\rho_2\theta^2\right]=0\,,
\end{split}
\end{equation} 
where we used~\eqref{eq:log-vanishes}, and in the second line we set $\hat\eta=0$ for simplicity, while in the third line we used the identification $2\rho_2=-\gamma^{b+3}\tilde \zeta$ and ignored the overall factor of $\gamma^b$. For the spatial projection, mutatis mutandis, we find 
\begin{equation}\label{eq:limitSprojection2}
\lim_{\gamma\to\infty} P_\alpha^\nu\nabla_\mu \hat T^{\mu\alpha}=\mathcal{E}a^\nu+v^\nu v^\mu\partial_\mu P_1+2v^\nu v^\mu\nabla_\mu(\rho_2\theta)+2\rho_2\theta a^\nu+\mathcal{O}(\gamma^{-2})=0\,,
\end{equation}
where again for simplicity we have set $\hat \eta=0$ and ignored the overall factor of $\gamma^b$. Contracting this last equation with $\tau_\nu$ and introducing it in \eqref{eq:limitSprojection2}, and using \eqref{eq:limituprojection2} in \eqref{eq:limitSprojection2}, leads to the two equations
\begin{equation}
\begin{split}
\mathcal{E}a^\nu+v^\nu\nabla_\mu(\mathcal{E}v^\mu)+2\rho_2\theta^2v^\nu+2v^\nu v^\mu\nabla_\mu(\rho_2\theta)+2\rho_2\theta a^\nu&=0 \,,\\
-\nabla_\mu\left(\mathcal{E}v^\mu\right)+\mathcal{E}a_\nu\tau^\nu-v^\mu\nabla_\mu(\rho_2\theta)+2\rho_2\theta a^\nu\tau_\nu-2\rho_2\theta^2&=0\,,
\end{split}
\end{equation}
in which the first corresponds to the combination of $\tau_\nu$ and $h_{\alpha\nu}$ projections of $\nabla_\mu T^{\mu\nu}=0$ and the second to the $\tau_\nu$ projection. It is interesting to note that from the limit we do not obtain the $v_\mu$ projection of $\nabla_\mu T^{\mu\nu}=0$ directly. However, in this case $v_\nu \nabla_\mu T^{\mu\nu}=-2\rho_3a^\mu a_\mu$, which vanishes due to the contraction of \eqref{eq:idealorderlimit-app} with $a_\nu$. This completes the analysis for the Landau frame.

\paragraph{Non-thermodynamic frame.} Moving on the the case of the non-thermodynamic frame, we use the redefinition freedom of relativistic hydrodynamics $T\to T+\delta T$ and $u^\mu\to u^\mu+\delta u^\mu$ to bring \eqref{eq:LandauST2} to
\begin{equation} \label{eq:LandauSTnon-thermo2}
    \hat T^{\mu\nu}_{(1)}=\hat\varsigma\hat\theta g^{\mu\nu}-\hat\eta \nabla^{(\mu}u^{\nu)}\,,    
\end{equation}
in which we have repeated the form in \eqref{eq:LandauSTnon-thermo} and where $\hat\varsigma=\tilde\zeta-s(\partial(Ts)/\partial T)^{-1}\tilde\zeta$. Here we also used the Euler relation $\varepsilon+\hat P=Ts$. Proceeding as in the core of the letter we can now scale the gradients of $u^\mu$ such that $\rho_1=\gamma^{b+}\hat\varsigma$ and $\eta=\gamma^{b+1}\hat\eta$ remain finite in the limit, leading to
\begin{equation}
\hat T^{\mu\nu}_{(1)}
\overset{\gamma\to \infty}{-\hspace{-0.2cm}\longrightarrow}\mathcal{E} v^\mu v^\nu+P g^{\mu\nu}+\rho_1\theta g^{\mu\nu}-\eta\sigma^{\mu\nu}\,,     
\end{equation}
which is the form written in \eqref{eq:stress1storder} with $\ell^\mu$ replaced by $v^\mu$. This case is slightly more interesting as we shall see. Considering for simplicity the case $\hat\eta=0$ we find 
\begin{equation} \label{eq:limituprojection3}
\begin{split}
u_\nu\nabla_\mu \hat T^{\mu\nu}&=-\nabla_\mu(w u^\mu)+u^\mu\partial_\mu \hat P+u_\mu\nabla^\mu\left(\hat\varsigma \theta\right)\\
&=\gamma U_\nu\nabla^\nu\left(\hat\varsigma \gamma^{b+1}\frac{\hat\theta}{\gamma}\right)+\frac{1}{\gamma}\left[-\nabla_\mu\left(w\gamma^2 U^\mu\right)+w\gamma^2U^\mu\partial_\mu \log  \gamma+\gamma^2 U^\mu \partial_\mu\left(\frac{P_1}{\gamma^2}\right)\right]=0\,,\\
&\overset{\gamma\to \infty}{-\hspace{-0.2cm}\longrightarrow}~\gamma v_\nu\nabla^\nu\left(\rho_1\theta\right)+\frac{1}{\gamma}\left[-\nabla_\mu\left(\mathcal{E}v^\mu\right)+v^\mu\partial_\mu P_1-2\rho_2\theta^2\right]=0\,,
\end{split}
\end{equation} 
where we used the identification $\rho_1=\hat\varsigma\gamma^{b+1}$ and~\eqref{eq:log-vanishes}. Similarly, we find
\begin{equation}\label{eq:limitSprojection3}
\lim_{\gamma\to\infty} P_\alpha^\nu\nabla_\mu \hat T^{\mu\alpha}=\gamma^2v^\nu v_\mu \nabla^\mu(\rho_1\theta)+\mathcal{E}a^\nu+v^\nu v^\mu\partial_\mu P_1+\nabla^\nu(\rho_1\theta)+\mathcal{O}(\gamma^{-2})=0\,.
\end{equation}
We see that, differently from the Landau frame, in this non-thermodynamic frame a leading order factor in $\gamma$ in \eqref{eq:limituprojection3} and $\gamma^2$ in \eqref{eq:limitSprojection3} appears and sets
\begin{equation} \label{eq:lprojection}
v^\mu\nabla_\mu(\rho_1\theta)=0\,,    
\end{equation}
which is precisely the projection $v_\nu\nabla_\mu T^{\mu\nu}=0$. The sub-leading terms then give the modified $\tau_\nu$ and $h_{\alpha\nu}$ projections as in earlier cases. We thus see that the null fluid equations are recovered from the limit and that Eq.~\eqref{eq:lprojection} gives dynamics to $\kappa$, once again showing that we find ourselves in the explicitly broken phase of null fluids. We also note that the appearance of leading terms in $\gamma$ in this non-thermodynamic frame suggests that one should consider the next order in the expansion in $1/\gamma^2$ in order to get an accurate form of the equations of motion. We leave a systematic expansion in $1/\gamma^2$ for future work.

\subsection{Limits of shear dispersion relations of relativistic fluids}
In this section we show how the limits of shear dispersion relations of timelike fluids, including gapped modes, agree with those in Section~\ref{app:modes} in the ultrarelativistic limit $\gamma\to\infty$. In \cite{Kovtun:2019hdm} all hydrodynamic and gapped modes were obtained for stress tensors written in thermodynamic frames. However, the stress tensor \eqref{eq:LandauSTnon-thermo2}, which we focused in the core of letter, is in a non-hydrodynamic frame. Therefore we must redo the mode analysis in 4 spacetime dimensions ($d=2$) by perturbing around an equilibrium state $T=T_0+\delta T$ and $u^\mu=u^{\mu}_0+\delta u^\mu$ where $T_0$ is a constant and $u^\mu_0=\gamma (1,0,0,v_0)$, in which $\gamma=(1-v_0^2)^{-1/2}$ is the Lorentz factor and $v_0$ the velocity along the $z$-direction. In the shear channel (i.e., $\delta T=0$) for the non-thermodynamic frame we find 
\begin{equation} \label{eq:shearmodes}
\begin{split}
\omega_{\text{sh},1}&=k_zv_0 +i\frac{\hat \eta}{2w \gamma}\left((k_z v_0)^2-k^2\right)+\mathcal{O}(k^3)\,,\\
\omega_{\text{sh},2}&=-i\frac{2w\gamma}{\hat\eta}-k_z v_0+i\frac{\hat\eta}{2w\gamma}\left(k^2-(k_z v_0)^2\right)+\mathcal{O}(k^3)\,,
\end{split}
\end{equation}
where $w=\varepsilon+\hat P$ is the enthalpy of the state. Interestingly, the expansion of the shear channel in small momenta coincides with the expansion in powers of $\gamma$ as $v_0\to1$ (for which the terms involving $k_z v_0$ must also be expanded and yield $k_z v_0\to k_z$ in the strict $\gamma\to\infty$ limit). Therefore, the small momenta expansion in the shear channel captures correctly the behaviour of the dispersion relations as $\gamma\to\infty$. The shear channel happens to coincide with the analysis of \cite{Kovtun:2019hdm} under certain identifications \footnote{In particular the coefficients $\theta_1,\theta_2$ and $\eta$ in \cite{Kovtun:2019hdm} are related to the coefficients here according to $\theta_1\to \hat\eta/2$, $\eta\to\hat\eta/2$ and $\theta_2=0$. The latter coefficient does not contribute to the shear channel in any case. For complete comparison with \cite{Kovtun:2019hdm} this frame also has $\varepsilon_1=\pi_1=0$, $\varepsilon_2=-\hat\varsigma$ and $\pi_2=\hat\varsigma-\eta/(d+1)$.}, in particular the mode $\omega_{\text{sh},2}$ had been given in \cite{Kovtun:2019hdm} but here we also explicitly included the $\mathcal{O}(k^2)$ correction. Given that we are considering fluids that obey the second law of thermodynamics $\hat\eta\ge0$ and $w\ge 0$ we note that there is no instability in this frame. In order to perform the limit $\gamma\to\infty$ we rescale $\omega\to\gamma^{b}\omega$ and $k\to\gamma^{b}k$, leading to
\begin{equation}
\begin{split}
\omega_{\text{sh},1}&=k_z v_0 +i\frac{\hat \eta \gamma^{b+1}}{2w \gamma^2}\left((k_z v_0)^2-k^2\right)+\mathcal{O}(k^3)\,, \\
\omega_{\text{sh},2}&=-i\frac{2w\gamma^2}{\gamma^{b+1}\hat\eta}-k_z v_0+i\frac{\gamma^{b+1}\hat\eta}{2w\gamma^2}\left(k^2-(k_z v_0)^2\right)+\mathcal{O}(k^3)\,.
\end{split}
\end{equation}
In the ultrarelativistic limit $\gamma\to\infty$, $v_0\to1$ \footnote{The limit $v_0\to-1$ can also be taken by changing the sign of some of the linear terms in $k_z$.}, choosing the $z$-direction such that $k_z v_0\to k_z$, while keeping $w\gamma^2=\mathcal{E}$ and $\eta=\gamma^{b+1}\hat\eta$ finite, we find
\begin{equation}
\begin{split}
\omega_{\text{sh},1}
&\overset{\gamma\to \infty}{-\hspace{-0.2cm}\longrightarrow}k_z-i\frac{\eta}{2\mathcal{E}}k_ik^i+\mathcal{O}(k^3)\,,\\
\omega_{\text{sh},2}
&\overset{\gamma\to \infty}{-\hspace{-0.2cm}\longrightarrow}-i\frac{2\mathcal{E}}{\eta}-k_z+i\frac{\eta}{2\mathcal{E}}k_i k^i+\mathcal{O}(k^3)\,.
\end{split}
\end{equation}
We see that these modes precisely coincide with those in \eqref{eq:modesgapped} and those in \eqref{eq:omega-pm-smallk} when $\rho_3=0$. This shows that the ultrarelativistic limit of the shear channel precisely coincides with part of the spectrum of null fluids. 

Focusing now in the other case in which the limit is taken directly in the Landau frame \eqref{eq:LandauST}, which is a thermodynamic frame, we can actually use the polynomials given in \cite{Kovtun:2019hdm} for both shear and sound channels in order to extract the modes. In \cite{Kovtun:2019hdm} the stress tensor parametrized in terms of the 6 transport coefficients $\vartheta,\pi_{1,2},\varepsilon_{1,2},\hat \eta$ and comparison with \eqref{eq:LandauST} we identify
\begin{equation} 
\label{eq:dictionarylandau}
\vartheta=0\,,\qquad \pi_1=0\,,\qquad \pi_2=\tilde\zeta\,,\qquad \varepsilon_{1,2}=0\,. 
\end{equation}
Written in this form, we can use the shear polynomial given in (4.3) of \cite{Kovtun:2019hdm}. In the Landau frame the modes read
\begin{equation} \label{eq:shearmodes2}
\begin{split}
\omega_{\text{sh},1}&=k_z v_0 +i\frac{\hat \eta}{2w \gamma}\left((k_z v_0)^2-k^2\right)+\mathcal{O}(k^3)\,,\\
\omega_{\text{sh},2}&=i\frac{2w}{\gamma v_0^2\hat\eta}+k_z v_0\left(\frac{2}{v_0}-v_0\right)+i\frac{\hat\eta}{2w\gamma}\left(k^2-(k_z v_0)^2\right)+\mathcal{O}(k^3)\,,
\end{split}
\end{equation}
where we have included $\mathcal{O}(k^2)$ corrections that were not explicitly given in \cite{Kovtun:2019hdm}. We also note that the first mode in \eqref{eq:shearmodes2} is the same as the first in \eqref{eq:shearmodes} but the other two are different in the different frames. Now taking the ultrarelativistic limit $\gamma\to\infty$, $v_0\to1$ along the $z$-direction and keeping $w_0\gamma^2$ as well as $2\rho_3=-\gamma^{b+3}\hat\eta$ finite we obtain
\begin{equation}
\begin{split}
\omega_{\text{sh},1}
&\overset{\gamma\to \infty}{-\hspace{-0.2cm}\longrightarrow}k_z+\;\mathcal O(k^3) \,,\\
\omega_{\text{sh},2}
&\overset{\gamma\to \infty}{-\hspace{-0.2cm}\longrightarrow}-i\frac{\mathcal{E}}{\rho_3} +k_z +\;\mathcal O(k^3)\,,   
\end{split}
\end{equation}
which agree with the modes in \eqref{eq:omega-pm-smallk} when $\eta=0$. As we noted earlier, when $\eta=0$ the modes \eqref{eq:omega-pm-smallk} are truncated at linear order in $k_z$, see \eqref{eq:modeseta0}, and thus predict that no additional corrections to $\omega_{\text{sh},1},\omega_{\text{sh},2}$ can appear in the limit. Indeed, one can check that the structure of the corrections to \eqref{eq:shearmodes2} is of the form $\sim (\hat\eta/w)^{n-1}\gamma^{-(n-1)}k^n$ for $n\ge2$, and therefore starting in Landau frame makes $\omega_{\text{sh},1}\big|_{\lim\gamma\to\infty}=k_z$ and $\omega_{\text{sh},2}\big|_{\lim\gamma\to\infty}=-i\mathcal{E}/\rho_3+k_z$ exact statements to all orders in $k$.

Since we assume the second law of thermodynamics $\hat \eta\ge0$ and $w\ge0$, we see that $\omega_{\text{sh},2}$ in the Landau frame is unstable. Consequently, since $\hat\eta\ge0$ and hence $\rho_3<0$, according to the criteria \eqref{eq:stabilitySSB}, in the limit the null fluid is unstable. This shows that the ultrarelativistic limit of the shear channel corresponds to part of the mode spectrum of null fluids, and that taking the limit of an unstable timelike fluid leads to an unstable null fluid.

\subsection{Limits of the sound channel of relativistic fluids}
The limits of the stress tensor taken in Section \ref{app:firstorderlimits} suggest that the full sound channel cannot be captured in the leading order ultrarelativistic limit since $P$ becomes constant. Here we show that all modes in the sound channel up to order $\mathcal{O}(\gamma^{-1})$ in the dispersion relations become the mode $\omega_0=k_z$ with the exact multiplicity of 3 as in the explicitly broken phase \eqref{eq:omega-pm-ESB} when $\chi_P=0$. However, contrary to the shear channel, $\mathcal{O}(\gamma^{-1})$ corrections to the dispersion relations are not accurately captured by the leading terms in the limits in Section \ref{app:firstorderlimits}. Starting with the non-thermodynamic frame, we compute the sound polynomial when $v_0=0$ leading to
\begin{equation}
F_{\text{sound}}(v_0=0,\omega,k)=i\frac{\hat\eta}{2}\omega^3-w\omega^2-i\left(\hat\gamma_s+\frac{\hat\eta v_s^2}{2}\right)\omega k^2+wv_s^2k^2\,,
\end{equation}
while for the Landau frame the sound polynomial when $v_0=0$ is quadratic and can be read directly from \cite{Kovtun:2019hdm} using \eqref{eq:dictionarylandau}. The boosted version of the polynomial (e.g. along $k_z$) can be obtained by sending $\omega\to\gamma(\omega-v_0 k_z)$ and $k_z\to\gamma(k_z-v_0\omega)$. In both frames we find the same pair of hydrodynamic modes, which can be written in the form
\begin{equation} \label{eq:soundmodes}
\omega_{\pm}=\Lambda_{\pm}(\bar k)k-i\Gamma_{\pm}(\bar k)k^2+\;\mathcal O(k^3)\,,\qquad \bar k=\frac{k_z}{k}\,,
\end{equation}
where we defined the phase and attenuation as
\begin{equation}
\Lambda_\pm(\bar k) =\frac{ v_0\,(1-v_s^2)\,\bar k
\ \pm\ \dfrac{v_s}{\gamma}\,
\sqrt{ \left( 1 - v_s^2 v_0^2 \right)\left( 1 - \bar k^2 \right)
       + \dfrac{\bar k^2}{\gamma^2} }
}
{ 1 - v_s^2 v_0^2 }\,,\qquad \Gamma_\pm(\bar k)=\frac{\hat \gamma_s}{2w}\gamma
\frac{\big(\Lambda_\pm(\bar k) - v_0\,\bar k\big)^2}{v_s^2}\,. 
\end{equation}
Here we introduced the speed of sound $v_s^2=\partial \hat P/\partial\varepsilon$ and defined $\hat\gamma_s=\hat\eta-(1+v_s^2)\hat\varsigma=\hat\zeta+2\hat\eta/3$. These expressions reduce to those of \cite{Kovtun:2019hdm} when $k_z=0$ or $k_i=0$. In order to take the ultrarelativistic limit we record the expansions of $\Lambda_\pm$ and $\Gamma_\pm$ as $v_0\to1$, in particular
\begin{equation}
\begin{split}
&\Lambda_\pm(\bar k)=\bar k +\mathcal{O}\left(\frac{1}{\gamma}\right)\,,\\
&\gamma\big(\Lambda_\pm(\bar k) - v_0\,\bar k\big)^2=\frac{v_s^2}{1-v_s^2}\frac{1-\bar k^2}{\gamma}+\mathcal{O}\left(\frac{1}{\gamma^2}\right)~\forall ~|\bar k|<1\,,\\
&\gamma\big(\Lambda_\pm(\bar k) - v_0\,\bar k\big)^2=\frac{v_s^2}{(1-v_s^2)^2\gamma^3}+\mathcal{O}\left(\frac{1}{\gamma^5}\right)~\iff ~|\bar k|=1\,.\\
\end{split}
\end{equation}
Differences in the two modes are sub-leading in the limit $v_0\to1$. Using these expansions the sound modes become
\begin{equation}
\begin{split}
\omega_\pm&=k_z-i\frac{\hat \gamma_s}{2w(1-v_s^2)\gamma}(1-\bar k^2)k^2+\mathcal{O}\left(k^3,\gamma^{-2}\right)~\forall~|\bar k|<1\,,\\
\omega_\pm&=k_z-i\frac{\hat \gamma_s}{2w(1-v_s^2)^2\gamma^3}k^2+\mathcal{O}\left(k^3,\gamma^{-5}\right)~\iff~|\bar k|=1\,.
\end{split}
\end{equation}
The ultrarelativistic limits, as in the shear channel, differ in both frames. In the non-thermodynamic frame, where $\gamma_s=\hat\gamma_s\gamma^{b+1}=\eta-(1+v_s^2)\rho_1$ is kept finite, we find 
\begin{equation} \label{eq:omegapmmodessound}
\begin{split}
\omega_\pm
&\overset{\gamma\to \infty}{-\hspace{-0.2cm}\longrightarrow}~k_z-i\frac{\gamma_s}{2\mathcal{E}(1-v_s^2)}k_ik^i+\mathcal{O}\left(k^3,\gamma^{-2}\right)~\forall~|\bar k|<1\,,\\
\omega_\pm
&\overset{\gamma\to \infty}{-\hspace{-0.2cm}\longrightarrow}~k_z+\mathcal{O}\left(k^3,\gamma^{-5}\right)~\iff~|\bar k|=1\,.
\end{split}
\end{equation}
It is clear that the case $|\bar k|<1$ exhibits a mode that is distinct from any of those found in \eqref{eq:omega-pm-ESB} for null fluids as it requires a new transport coefficient $\gamma_s$ and knowledge of the speed of sound. In particular, the terms involving $k^i k_i$ in \eqref{eq:omega-pm-ESB} are proportional to $\eta$ rather than $\gamma_s$. We note that the resulting modes are stable since $\mathcal{E},\gamma_s\ge0$ and $v_s<1$. On the other hand, if the limit is taken starting in the Landau frame and keeping $\gamma_s=\hat\gamma_s\gamma^{b+3}=-(\rho_2+\rho_3)/2$ finite, we find $\omega_\pm=\omega_0=k_z+\mathcal{O}(k^3,\gamma^{-2})$ for all $|\bar k|\le1$ - a result that seems to hold to arbitrary high order (we checked up to $\mathcal{O}(k^7)$). In turn, the gapped mode in the sound channel is different in both frames. In the non-thermodynamic frame the gapped mode takes the form
\begin{equation}
\omega_{\text{gap}}=-i\frac{2w(1-v_0^2v_s^2)}{\gamma(\hat\eta-v_0^2(2\hat\gamma_s+\hat\eta v_s^2))}+A_1 k_z v_0+ A_2 (k_z v_0)^2+B_2 k_i k^i+\mathcal{O}(k^3)\,,
\end{equation}
where $A_1$ and $B_2$ are given by
\begin{equation}
\begin{split}
   A_1&= v_0-\frac{4v_0\hat\gamma_s}{\gamma^2(1-v_0^2v_s^2)(\hat\eta-v_0^2(2\hat\gamma_s+\eta v_s^2))}\,,\qquad A_2=i\frac{\hat\gamma_s(1+v_0^3v_s^2)}{\gamma^3w(1-v_0^2v_s^2)^3}\,,\qquad B_2=\frac{i}{w\gamma}\frac{\hat\gamma_s}{(1-v_s^2v_0^2)^2}\,.
\end{split}
\end{equation}
When $v_0=0$ the mode is stable but for non-zero $v_0$ it develops an instability at a critical value of $v_0$ since $\hat\eta, \hat \zeta\ge0$. In the ultrarelativistic limit we obtain
\begin{equation} \label{eq:modegappedsoundnonthermo}
\omega_{\text{gap}}
\overset{\gamma\to \infty}{-\hspace{-0.2cm}\longrightarrow}~k_z+i\frac{\gamma_s}{\mathcal{E}(1-v_s^2)^2}k_i k^i+\mathcal{O}(k^3)~\iff~\text{non-thermodynamic frame}\,,
\end{equation}
where we used the fact that $A_1\to1$, $A_2\sim\hat\eta/(w\gamma^3)$ and $B_2\sim\hat\eta/(w\gamma)$ as $\gamma\to\infty$.

On the other hand, in the Landau frame, the gapped mode is only visible at non-zero $v_0$ and reads
\begin{align}
\omega_{\text{gap}}
&= i\frac{w(1 - v_0^2 v_s^2)}{\hat\gamma_s\,\gamma\,v_0^2}
+\frac{\,v_0^4 v_s^2 + v_0^2 - 2\,}{\,v_0\,(v_0^2 v_s^2 - 1)\,}k_z v_0  + i\,\frac{\hat\gamma_s}{w\gamma\,\big(1 - v_0^2 v_s^2\big)^2}\,k_i k^i
\;+\; i\,\frac{\hat \gamma_s\,(1+3 v_0^2 v_s^2)}{w\gamma^3\big(1 - v_0^2 v_s^2\big)^3}\,(k_z v_0)^2
\;+\;\mathcal O(k^3)\,.
\end{align}
This gapped mode in the Landau frame is unstable for any non-zero $v_0$ since $\hat\eta, \hat \zeta\ge0$ for fluids obeying the second law constraints and its ultrarelativistic limit yields
\begin{equation}
\begin{split} \label{eq:gapsoundLandau}
\omega_{\text{gap}}
&\overset{\gamma\to \infty}{-\hspace{-0.2cm}\longrightarrow}~i\frac{\mathcal{E}(1 - v_s^2)}{\gamma_s}+k_z+\;\mathcal O(k^3)~\iff~\text{Landau frame}\,.
\end{split}    
\end{equation}
The gapped modes in the ultrarelativistic limit in both frames are also different, and appear to exhibit instabilities. When the limit is taken starting from the Landau frame, the instability in the corresponding null fluid is inherited from the instability in the Landau frame. On the other hand, the instability appearing in the limit of the non-thermodynamic frame is not arising from an instability in the original timelike fluid since the gapped mode became gapless in limit. This instability is likely appearing due to the absence of additional $\mathcal{O}(\gamma^{-1})$ corrections that we did not take into account when taking the limit of these dispersion relations, and which we discuss below in more detail.

As we mentioned in the beginning of this section, all modes in both frames, including the gapped modes, reduce to $\omega_0=k_z$ when ignoring $\mathcal{O}(\gamma^{-1})$ corrections. Given that there are two sound modes and one gapped mode, the multiplicity of $\omega_0$ is 3 and matches the multiplicity of the $\omega_0$ modes in the explicitly broken phase given in section \ref{app:modes} when $\chi_P=0$. The corrections of order $\mathcal{O}(\gamma^{-1})$ that we obtained in this section do not agree with the modes in section \eqref{app:modes} for any $\chi_P$. In section \ref{app:firstorderlimits} it was already noted that $\mathcal{O}(\gamma^{-1})$ corrections to the limit of the stress tensor can be important for a perfect match with the equations of motion. Together, these results suggest that $\mathcal{O}(\gamma^{-1})$ corrections are needed in order to understand the ultrarelativistic limit of the sound channel or that an additional appropriate expansion of the sound channel is needed. The naive limits taken in this section will likely be refined when a $\mathcal{O}(\gamma^{-1})$ expansion is performed. Therefore, the specific corrections obtained here should be viewed only as part of the contributions that are expected to appear at $\mathcal{O}(\gamma^{-1})$. We leave these questions for future work.

\subsection{Limits of scalar field Lagrangians and null matter}
It is well known that the energy-momentum tensors of many classes of scalar field theories can be recast in the form of a fluid stress tensor upon certain identifications (see, e.g.,~\cite{Faraoni:2018fil,Faraoni:2018qdr}). It has also been shown that simple Langrangians for the scalar field $\Phi$ can lead to null dust, for which the stress tensor takes the form of $\eqref{eq:idealstress}$ but with $P=0$ \cite{Faraoni:2018fil}. Here we show that these simple Lagrangians can also accommodate null matter, where $P$ does not necessarily vanish. We also show that when coupling to dynamical gravity in the context of Brans--Dicke theory, the stress tensor can acquire higher-derivative terms that match \eqref{eq:stress1storder}.

\paragraph{Scalar field Lagrangian.} We consider the following scalar field theory with a slightly unusual kinetic term
\begin{equation} \label{eq:scalarLagrangian}
S_\Phi=\int d^Dx\left(-\frac{f(\Phi)}{2}\nabla_\mu \Phi\nabla^\mu\Phi-V(\Phi)\right)\,,
\end{equation}
where $f(\Phi)$ and $V(\Phi)$ are arbitrary functions of $\Phi$ and $D=d+2$. We can straightforwardly compute the stress tensor by variation:
\begin{equation}\label{eq:scalarstress}
T^{\mu\nu}=-\frac{2}{\sqrt{-g}}\frac{\delta S_\Phi}{\delta g_{\mu\nu}}=f(\Phi) \nabla^\mu \Phi\nabla^\nu \Phi-\frac{f(\Phi)}{2}g^{\mu\nu}\nabla_\alpha \Phi\nabla^\alpha\Phi-g^{\mu\nu}V(\Phi)\,.    
\end{equation}
In the timelike case, one assumes that $\nabla_\alpha \Phi\nabla^\alpha\Phi<0$, thus bringing the stress tensor \eqref{eq:scalarstress} to the perfect fluid form \eqref{eq:stressidealtimelike} with the identifications 
\begin{equation}\label{eq:timescalar}
u^\mu=\frac{\partial^\mu\Phi}{\sqrt{-\nabla_\alpha \Phi\nabla^\alpha\Phi}}\,,\qquad \varepsilon+\hat P=-f(\Phi)\nabla_\alpha \Phi\nabla^\alpha\Phi\,,\qquad \hat P=-\frac{f(\Phi)}{2}\nabla_\alpha \Phi\nabla^\alpha\Phi-V(\Phi)\,.  
\end{equation}
In the null case, we can proceed as in \cite{Faraoni:2018fil} and assume that instead $\nabla_\alpha \Phi\nabla^\alpha\Phi=0$. We thus identify
\begin{equation} \label{eq:nullscalar}
v^\mu =\partial^\mu\Phi\,,\qquad \mathcal{E}=f(\Phi)\,,\qquad P=-V(\Phi)\,,    
\end{equation}
thus bringing \eqref{eq:scalarstress} to the ideal null form \eqref{eq:idealstress}. We can also obtain this form directly from the limit of the timelike case \eqref{eq:timescalar}; in particular, defining $U^\mu=\partial^\mu \Phi$ and $\gamma^{-1}=\sqrt{-\nabla_\alpha\Phi \nabla^\alpha \Phi}$, one finds 
\begin{equation}
\begin{split}
u^{\mu}u^{\nu}(\varepsilon+\hat P)=\frac{-U^{\mu}U^{\nu}f(\Phi)|\gamma|^{2}}{\gamma^{2}}=U^{\mu}U^{\nu}f(\Phi)\to v^{\mu}v^{\nu}f(\Phi)\,,~~~\hat{P}=-\frac{f(\Phi)}{2\gamma^2} - V(\Phi)  \to -V(\Phi)\,, 
\end{split}
\end{equation}
thus again leading to \eqref{eq:nullscalar}. We see that in this case we do not need to send a specific parameter, such as the temperature in the case of fluids, to a limiting value since the factors of $\gamma$ precisely cancel each other in the combination $u^{\mu}u^{\nu}(\varepsilon+\hat P)$. Differently from \cite{Faraoni:2018fil}, which focused on null dust ($P=0$), here we have allowed for a non-zero pressure and showed that the stress tensor takes the form of perfect null matter. One can show that the equations of motion \eqref{eq:nullequations} follow for the stress tensor \eqref{eq:scalarstress} due to diffeomorphism symmetry of the Lagrangian and the equation of motion for the scalar field 
\begin{equation}
\frac{\delta S_\Phi}{\delta\Phi}=f(\Phi)\square\Phi+\frac{1}{2}\frac{\partial f(\Phi)}{\partial \Phi}\nabla_\alpha \Phi\nabla^\alpha\Phi-\frac{\partial V(\Phi)}{\partial \Phi}=0\,,
\end{equation}
where $\square =\nabla_\mu \nabla^\mu$. While this identification works to what concerns the form of the stress tensor it is important to note that there is only one degree of freedom at the end, namely $\Phi$, instead of both $v_\mu$ and $\kappa$ that we introduced in \eqref{eq:idealstress}. To make the mapping precise we would need to assume that $v_\mu= \partial_\mu\Phi$ and $\kappa=\Phi$, which in general leads to a different low energy spectrum than that presented in Section \ref{app:modes}; see below for details.

\paragraph{Brans-Dicke theory.} Finding Lagrangians that can capture generic high-order corrections typically requires working with Schwinger--Keldysh effective field theory. Here instead we mimick such effects by minimally coupling the scalar field theory \eqref{eq:scalarLagrangian} to dynamical gravity. We thus consider the following Brans--Dicke theory action
\begin{equation} \label{eq:lagrangianBD}
S_{\text{BD}}=\int d^Dx\left(R\Phi-\frac{f(\Phi)}{2}\nabla_\mu \Phi\nabla^\mu\Phi-V(\Phi)\right)\,,    
\end{equation}
where $R$ is the Ricci scalar and $f(\Phi)$ has now the interpretation of the Brans--Dicke coupling. The effective stress tensor for the scalar field that can be derived from \eqref{eq:lagrangianBD} once using the Einstein equations obtained from $\delta S_{\text{BD}}/\delta g_{\mu\nu}=0$ takes the form \cite{Faraoni:2018qdr}
\begin{equation} \label{eq:stressBD}
T^{\mu\nu}_{\text{BD}}=-\frac{2}{\sqrt{-g}}\frac{\delta S_{\text{BD}}}{\delta g_{\mu\nu}}=\frac{f(\Phi)}{2\Phi}\left(\nabla^\mu\Phi\nabla^\nu\Phi-\frac{1}{2}g^{\mu\nu}\nabla_\alpha\Phi\nabla^\alpha\Phi\right)+\frac{1}{\Phi}\left(\nabla^\mu\nabla^\mu\Phi-g^{\mu\nu}\square \Phi\right)-\frac{V(\Phi)}{2\Phi}g^{\mu\nu}\,,   
\end{equation}
and is covariantly conserved $\nabla_\mu T^{\mu\nu}_{\text{BD}}=0$. It can be shown that in general the stress tensor \eqref{eq:stressBD} can be put into the form of an anisotropic fluid and with first order gradient corrections due to terms involving two derivatives in \eqref{eq:stressBD}. We now focus on the null case $\nabla_\alpha \Phi\nabla^\alpha\Phi=0$ and rewrite \eqref{eq:stressBD} in the general form \eqref{eq:stresscorrectionESB}, where 
\begin{equation}
\mathcal{E}=\frac{f(\Phi)}{2\Phi}\,,\qquad v^\mu=\partial^\mu\Phi\,,\qquad P=-\frac{V(\Phi)}{2\Phi}\,,\qquad \rho_1=-\frac{(d+1)}{(d+2)\Phi}\,,\qquad \eta=-\frac{1}{\Phi}\,. 
\end{equation}
The coefficients appearing here were introduced in \eqref{eq:generalframe}. If we were to treat the gradients perturbatively then we could redefine $v^\mu$ as to remove $\rho_1$ as explained in the main text. We have thus shown that scalar field theories with null gradients are one interesting example of null matter and that the effective theory we developed in this work is able to capture the form of their respective stress tensors. We note that making the identification $v_\mu=\partial_\mu \Phi$ and $\kappa=\Phi$ and computing the modes leads to
\begin{equation}
 \omega^\Phi_\pm=\pm\sqrt{k^2+m^2}\,, \qquad m^2=\frac{\chi_P}{\eta-\rho_1}\,,   
\end{equation}
taking a form similar to excitations of a massive scalar field and where all quantities are evaluated in the equilibrium state $\kappa=\kappa_0=\Phi_0$. In this specific example $m^2<0$ if $\chi_P>0$ and hence the modes  $\omega^\Phi_\pm$ are purely imaginary at low $k$ leading to an instability. Otherwise, if $\chi_P<0$, the modes are stable and gapped.

\end{document}